\documentclass[amsmath, pra, twocolumn, showpacs, floatfix]{revtex4-1}
\usepackage{graphicx}
\usepackage{dsfont}

\begin{document}   

\title{Nanofiber-mediated chiral radiative coupling between two atoms}
 
\author{Fam Le Kien}

\affiliation{Quantum Systems Unit, Okinawa Institute of Science and Technology Graduate University, Onna, Okinawa 904-0495, Japan}

\author{A. Rauschenbeutel} 

\affiliation{Vienna Center for Quantum Science and Technology, Institute of Atomic and Subatomic Physics, Vienna University of Technology, Stadionallee 2, 1020 Vienna, Austria}

\date{\today}

\begin{abstract}
We investigate the radiative coupling between two two-level atoms with arbitrarily polarized dipoles in the vicinity of a nanofiber. We present a systematic derivation for the master equation, the single- and cross-atom decay coefficients, and the dipole-dipole interaction coefficients for the atoms interacting with the vacuum of the field in the guided and radiation modes of the nanofiber. We study numerically the case where the atomic dipoles are circularly polarized. In this case, the rate of emission depends on the propagation direction, that is, the radiative interaction between the atoms is chiral. We examine the time evolution of the atoms for different initial states. We calculate the fluxes and mean numbers of photons spontaneously emitted into guided modes in the positive and negative directions of the fiber axis. We show that the chiral radiative coupling modifies the collective emission of the atoms. We observe that the modifications strongly depend on the initial state of the atomic system, the radiative transfer direction, the distance between the atoms, and the distance from the atoms to the fiber surface. 
\end{abstract}

\maketitle

\section{Introduction}

Radiative coupling between two atoms (or molecules) has been a topic of great interest for the past several decades. The range and strength of the coupling
can be enhanced  by means of ``dressing'' the environment. A typical nonradiative F\"{o}rster energy transfer 
range of $\leq10$ nm was surpassed by use of localized plasmons, whispering 
gallery modes, or microcavities \cite{Gersten,Folan,Arnold,Agarwal98,Hopmeier,Barnes,Hartman,Basko,Dung,Benson2006,Meixner2014,Tsai2014}. 
Very fast (on the picosecond time scale) energy transfer was 
recorded in systems with quantum dots \cite{Crooker} and strongly bound excitons \cite{Kozawa2016}. 
Plasmon-assisted communication was demonstrated between donor-acceptor pairs across 120-nm-thick metal films \cite{Andrew} and between fluorophores on top of a silver film over distances up to 7 $\mu$m \cite{Bouchet2016}. The recent directions of research on 
dipole-dipole interaction now encompass areas of few-atom spectroscopy \cite{Hettich,Zhang2016}, near-field 
optics \cite{Pohl}, and subwavelength-resolution nano-optics \cite{nano-optics}. 
The effects of a nanosphere on the dipole-dipole interaction have been studied \cite{Klimov98,Dung2}. 
A form of ``telegraphy'' on a dielectric microplanet has been proposed \cite{Arnold}. It is clear that 
the range of such ``telegraphy'' can be increased arbitrarily if one uses nanofibers. 

The effects of a nanofiber on spontaneous emission of a two-level atom \cite{Jhe,Klimov}, a multilevel atom \cite{cesium decay}, and two two-level atoms \cite{twoatoms} have been studied. It has been shown that spontaneous emission and scattering from an atom with a circular dipole in front of a nanofiber can be asymmetric with respect to the opposite axial propagation directions \cite{Mitsch14b,Petersen14,Fam14,AtomArray,Scheel15,Sayrin15b}. These directional effects are the signatures of spin-orbit coupling of light \cite{Zeldovich,Bliokh review,Bliokh review2015,Bliokh2014,Bliokh2015} carrying transverse spin angular momentum \cite{Bliokh2014,Banzer review2015}. They are due to the existence of a nonzero longitudinal component of the nanofiber guided field, which oscillates in phase quadrature with respect to the radial transverse component. The possibility of directional emission from an atom into propagating radiation modes of a nanofiber and the possibility of generation of a lateral force on the atom have been pointed out \cite{Scheel15}. The direction-dependent emission and absorption of photons lead to chiral quantum optics \cite{Lodahl2016}.
It has been shown that substantial coupling between two atoms can survive over long interatomic distances due to guided modes \cite{twoatoms}. 
The chiral coupling between atoms has been studied in the framework of one-dimensional waveguide bath models \cite{Stannigel2012,Ramos2014,Zoller2015,Gonzalez-Ballestero2015,Ramos2016,Vermersch2016,Eldredge2016}, where radiation modes were completely  \cite{Stannigel2012,Ramos2014,Zoller2015,Ramos2016,Vermersch2016} or partially  \cite{Gonzalez-Ballestero2015,Eldredge2016} neglected. 
In closely related studies, the chiral effect in spontaneous emission of a single atom \cite{Fam16}
and the radiative transfer between two atoms \cite{Yuan2016} in front of a dielectric surface have also been investigated.

In this paper, we study radiative coupling between two two-level atoms with arbitrarily polarized dipoles
in the vicinity of a nanofiber. Unlike Ref.~\cite{twoatoms}, our treatment incorporates rotating induced dipoles.
In addition, our treatment is more general than the previous studies \cite{Stannigel2012,Ramos2014,Zoller2015,Gonzalez-Ballestero2015,Vermersch2016} in the sense that we use a three-dimensional fiber model and take into account the effects of radiation modes on the decay rates and the dipole-dipole interaction coefficients. 
We focus on the case where the atomic dipoles are circularly polarized and, consequently, the rate of emission depends on the propagation direction and the radiative interaction between the atoms is chiral. In order to get insight into this chiral coupling, we look at the decay behavior of the atoms as well as the fluxes and numbers of photons emitted into guided modes.

The paper is organized as follows. In Sec.\ \ref{sec:model} we describe the model of two two-level atoms with arbitrarily polarized dipoles in the vicinity of a nanofiber. In Sec.\ \ref{sec:analytical} we derive the basic equations for the interaction between the atoms and the field
in guided and radiation modes. In Sec.\ \ref{sec:numerical} we present the results of numerical calculations. Our conclusions are given in Sec.~\ref{sec:summary}.

\section{Model}
\label{sec:model}

\subsection{Quantization of the field around a nanofiber}

\begin{figure}[tbh]
\begin{center}
  \includegraphics{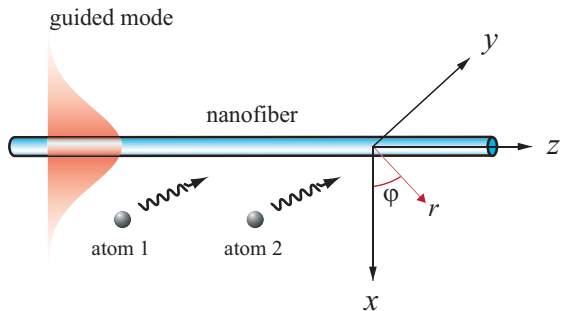}
 \end{center}
\caption{Two two-level atoms in the vicinity of a nanofiber.
}
\label{fig1}
\end{figure}

We consider a fiber that has a cylindrical silica core of radius $a$ and refractive index $n_1>1$ and an infinite vacuum clad of refractive index $n_2=1$ (see Fig.~\ref{fig1}). We use the Cartesian coordinates $\{x,y,z\}$ and the cylindrical coordinates $\{r,\varphi,z\}$ with $z$ being the fiber axis. In view of the very low losses of silica in the wavelength range of interest, we neglect material absorption. 

The continuum field quantization follows the procedures presented in Ref. \cite{Loudon}.
In the presence of the nanofiber, the positive-frequency part $\mathbf{E}^{(+)}$ of the electric component of the field can be
decomposed into the contributions $\mathbf{E}^{(+)}_{\mathrm{gyd}}$ and $\mathbf{E}^{(+)}_{\mathrm{rad}}$ from guided and radiation modes, respectively, as 
\begin{equation}\label{d1}
\mathbf{E}^{(+)}=\mathbf{E}^{(+)}_{\mathrm{gyd}}+\mathbf{E}^{(+)}_{\mathrm{rad}}.
\end{equation}

Regarding guided modes, we assume that the single-mode condition \cite{fiber books} 
is satisfied for a finite bandwidth of the field frequency $\omega$ around a characteristic atomic transition frequency $\omega_0$. 
In this case, the nanofiber supports only the hybrid fundamental modes HE$_{11}$ corresponding to the wavelength $\lambda=2\pi c/\omega$ \cite{fiber books}.
We label each guided mode by an index $\mu=(\omega fl)$, 
where $f=+,-$ denotes the forward or backward propagation direction,
and $l=+,-$ denotes the counterclockwise or clockwise polarization. When we quantize the field in the guided modes, we obtain the following expression 
for $\mathbf{E}^{(+)}_{\mathrm{gyd}}$ in the interaction picture:
\begin{equation}
\mathbf{E}^{(+)}_{\mathrm{gyd}}=i\int _0^{\infty}d\omega\sum_{fl}\sqrt{\frac{\hbar\omega\beta'}{4\pi\epsilon_0}}
\;a_{\mu}\mathbf{e}^{(\mu)}e^{-i(\omega t-f\beta z-l\varphi)}.
\label{2}
\end{equation}
Here $\beta$ is the longitudinal propagation constant, $\beta'$ is the derivative of $\beta$
with respect to $\omega$, $a_{\mu}$ is the respective photon annihilation operator, and $\mathbf{e}^{(\mu)}=\mathbf{e}^{(\mu)}(r,\varphi)$ is 
the electric-field profile function of the guided mode $\mu$ in the classical problem. The constant $\beta$ is determined by the
fiber eigenvalue equation (\ref{g1}).
The operators $a_{\mu}$ and $a_{\mu}^\dagger$ satisfy the continuous-mode bosonic commutation rules
$[a_{\mu},a_{\mu'}^\dagger]=\delta(\omega-\omega')\delta_{ff'}\delta_{ll'}$. 
The normalization of $\mathbf{e}^{(\mu)}$ is given by
\begin{equation}
\int _{0}^{2\pi}d\varphi\int _{0}^{\infty}n_{\mathrm{ref}}^2\,|\mathbf{e}^{(\mu)}|^2r\,dr=1.
\label{3}
\end{equation}
Here $n_{\mathrm{ref}}(r)=n_1$ for $r<a$, and $n_{\mathrm{ref}}(r)=n_2$ for $r>a$.
The explicit expression for the guided mode function $\mathbf{e}^{(\mu)}$ is given
in Appendix \ref{sec:guided}.

Unlike the case of guided modes, in the case of radiation modes,
the longitudinal propagation constant $\beta$ for each value of $\omega$ can vary continuously, from $-k$ to $k$, where $k=\omega/c$ is the wavelength of light in free space. We label each radiation mode by an index $\nu=(\omega\beta ml)$, where 
$m=0,\pm1,\pm2,\dots$ is the mode order and $l=\pm$ is the mode polarization. 
When we quantize the field in the radiation modes, we obtain the following expression 
for $\mathbf{E}^{(+)}_{\mathrm{rad}}$ in the interaction picture:
\begin{eqnarray}\label{d8}
\mathbf{E}^{(+)}_{\mathrm{rad}}&=&i\int _0^{\infty}d\omega\int _{-k}^{k}d\beta\sum_{ml}
\sqrt{\frac{\hbar\omega}{4\pi\epsilon_0}}\;a_{\nu}\mathbf{e}^{(\nu)}
\nonumber\\&&\mbox{}\times
e^{-i(\omega t-\beta z-m\varphi)}.
\end{eqnarray}
Here, $a_{\nu}$ is the respective photon annihilation 
operator, and $\mathbf{e}^{(\nu)}=\mathbf{e}^{(\nu)}(r,\varphi)$ is the electric-field profile function of the radiation mode $\nu$ 
in the classical problem. 
The operators 
$a_{\nu}$ and $a_{\nu}^\dagger$ satisfy the continuous-mode bosonic commutation rules
$[a_{\nu},a_{\nu'}^\dagger]=\delta(\omega-\omega')\delta(\beta-\beta')
\delta_{mm'}\delta_{ll'}$. 
The normalization of $\mathbf{e}^{(\nu)}$ is given by
\begin{eqnarray}\label{d9}
&&\int _0^{2\pi}d\varphi\int _{0}^{\infty}n_{\mathrm{ref}}^2
\left[\mathbf{e}^{(\nu)}\mathbf{e}^{(\nu')*}\right]_{\beta=\beta',m=m',l=l'}
r\,dr\nonumber\\&&
=\delta(\omega-\omega').
\end{eqnarray}
The explicit expression for the radiation mode function $\mathbf{e}^{(\nu)}$ is given
in Appendix \ref{sec:radiation}.

\subsection{Two atoms interacting with the field}

Consider two two-level atoms with the identical transition frequency $\omega_0$. We label the atoms by the index $j=1,2$.
The atoms are located at points $\mathbf{R_1}\equiv \{r_1,\varphi_1,z_1\}$ and $\mathbf{R_2}\equiv\{r_2,\varphi_2,z_2\}$ (see Fig.~\ref{fig1}). 
In the interaction picture, the electric dipole of atom $j$ is given by
$\mathbf{D}_j=\mathbf{d}_{j}^*\sigma_{j}e^{-i\omega_0t}+\mathbf{d}_{j}\sigma_{j}^\dagger e^{i\omega_0t}$.
Here, the operators $\sigma_{j}=|-\rangle_j{}_j\langle +|$ and 
$\sigma_{j}^\dagger=|+\rangle_j{}_j\langle -|$ 
describe respectively the downward and upward transitions of atom $j$, and 
$\mathbf{d}_{j}$ is the corresponding dipole matrix element. The notations $|+\rangle_j$ and $|-\rangle_j$ 
stand for the upper and lower states, respectively, of atom $j$. 
In general, the dipole matrix element $\mathbf{d}_{j}$ can be a complex vector.
The basis states of the two-atom system can be written as $|s_1s_2\rangle=|s_1\rangle_1\otimes|s_2\rangle_2$,
where $s_1,s_2=\pm$.

For brevity, we use the index $\alpha=\mu,\nu$ as a common label for the guided modes $\mu$ and the radiation modes $\nu$. In addition, we use the notation
$\sum_{\alpha}=\sum_{\mu}+\sum_{\nu}$, where  
$\sum_{\mu}=\int_0^{\infty}d\omega\sum_{fl}$ and $\sum_{\nu}=\int_0^{\infty}d\omega\int_{-k}^{k}d\beta\sum_{ml}$ are generalized
summations over guided and radiation modes, respectively.
In the interaction picture, the Hamiltonian for the atom-field interaction in the dipole approximation can be written as  
\begin{equation}\label{9}
\begin{split}
H_{\mathrm{int}}&=-i\hbar\sum_{\alpha j}(G_{\alpha j}\sigma_{j}^{\dagger} a_{\alpha}e^{-i(\omega-\omega_0)t}-\mbox{H.c.})\\
&\quad
-i\hbar\sum_{\alpha j}(\tilde{G}_{\alpha j}\sigma_{j} a_{\alpha}e^{-i(\omega+\omega_0)t}-\mbox{H.c.}).\\
\end{split}
\end{equation}
Here, the coefficient $G_{\alpha j}$ characterizes the coupling of atom $j$  with
mode $\alpha$ via the co-rotating terms $\sigma_j^\dagger a_\alpha$ and $\sigma_j a_\alpha^\dagger$.  
The expressions for $G_{\alpha j}$ with $\alpha=\mu,\nu$ are 
\begin{equation}\label{8}
\begin{split}
G_{\mu j}&=\sqrt{\frac{\omega\beta'}{4\pi\hbar\epsilon_0}}\;
\big[\mathbf{d}_{j}\cdot\mathbf{e}^{(\mu)}(r_j,\varphi_j)\big]e^{i(f\beta z_j+l\varphi_j)},\\
G_{\nu j}&=\sqrt{\frac{\omega}{4\pi\hbar\epsilon_0}}\;
\big[\mathbf{d}_{j}\cdot\mathbf{e}^{(\nu)}(r_j,\varphi_j)\big]e^{i(\beta z_j+m\varphi_j)}.
\end{split}
\end{equation}
The coefficient $\tilde{G}_{\alpha j}$ describes the coupling of atom $j$ with
mode $\alpha$ via the counter-rotating terms $\sigma_j a_\alpha$ and $\sigma_j^\dagger a_\alpha^\dagger$. 
The expressions for $\tilde{G}_{\alpha j}$ with $\alpha=\mu,\nu$ are obtained from Eqs.~\eqref{8} by replacing
the dipole matrix element $\mathbf{d}_{j}$ with its complex conjugate $\mathbf{d}_{j}^*$, that is,
\begin{equation}\label{8a}
\begin{split}
\tilde{G}_{\mu j}&=\sqrt{\frac{\omega\beta'}{4\pi\hbar\epsilon_0}}\;
\big[\mathbf{d}^*_{j}\cdot\mathbf{e}^{(\mu)}(r_j,\varphi_j)\big]e^{i(f\beta z_j+l\varphi_j)},\\
\tilde{G}_{\nu j}&=\sqrt{\frac{\omega}{4\pi\hbar\epsilon_0}}\;
\big[\mathbf{d}^*_{j}\cdot\mathbf{e}^{(\nu)}(r_j,\varphi_j)\big]e^{i(\beta z_j+m\varphi_j)}.
\end{split}
\end{equation}

\section{Basic equation}
\label{sec:analytical}

\subsection{Master equation for the atoms}

We call $\mathcal{O}$ an arbitrary atomic operator. The Heisenberg equation for this operator is
\begin{equation}\label{12}
\begin{split}
\dot{\mathcal{O}}&=\sum_{\alpha j}(G_{\alpha j}
[\sigma_{j}^{\dagger},\mathcal{O}] a_{\alpha}
e^{-i(\omega-\omega_0)t}\\
&\quad
+\tilde{G}_{\alpha j}
[\sigma_{j},\mathcal{O}] a_{\alpha}
e^{-i(\omega+\omega_0)t}\\
&\quad+G_{\alpha j}^{*}a_{\alpha}^{\dagger}[\mathcal{O},\sigma_{j}]
e^{i(\omega-\omega_0)t}\\
&\quad
+\tilde{G}_{\alpha j}^{*}a_{\alpha}^{\dagger}[\mathcal{O},\sigma_{j}^\dagger]
e^{i(\omega+\omega_0)t}).
\end{split}
\end{equation}
The Heisenberg equation for the photon annihilation operator $a_{\alpha}$ is
\begin{equation}\label{10}
\dot{a}_{\alpha}=\sum_{j}G_{\alpha j}^*\sigma_{j}e^{i(\omega-\omega_0)t}
+\sum_{j}\tilde{G}_{\alpha j}^*\sigma_{j}^\dagger e^{i(\omega+\omega_0)t}.
\end{equation}
We integrate Eq. (\ref{10}). Then, we obtain
\begin{equation}\label{11}
\begin{split}
a_{\alpha}(t)&=a_{\alpha}(t_0)+\sum_{j}G_{\alpha j}^*\int\limits _{t_0}^t dt'\,
\sigma_{j}(t')e^{i(\omega-\omega_0)t'}\\
&\quad+\sum_{j}\tilde{G}_{\alpha j}^*\int\limits _{t_0}^t dt'\,
\sigma_{j}^\dagger (t')e^{i(\omega+\omega_0)t'},
\end{split}
\end{equation}
where $t_0$ is the initial time.

We consider the situation where the field is initially in the vacuum state. 
We assume that the evolution time $t-t_0$ and the characteristic atomic lifetime $\tau_a$ are 
large as compared to the optical period $2\pi/\omega_0$ and 
the light propagation time $|\mathbf{R_2}-\mathbf{R_1}|/c$ 
between the two atoms. When the continuum of the guided and radiation modes is regular and broadband around the atomic frequency,
the effect of the retardation is concealed \cite{Ujihara}, and
the Markov approximation $\sigma_{j}(t')=\sigma_{j}(t)$ can be applied to describe the back
action of the second and third terms in Eq. (\ref{11}) on the atom. 
Under the condition $t-t_0\gg 2\pi/\omega_0$, 
we calculate the integrals with respect to $t'$ in the limit $t-t_0\to\infty$.
Then, Eq.~\eqref{11} yields
\begin{equation}\label{11a}
\begin{split}
&a_{\alpha}(t)=a_{\alpha}(t_0)\\
&\quad+\sum_{j}G_{\alpha j}^*\sigma_{j}(t)e^{i(\omega-\omega_0)t}
\bigg[\pi\delta(\omega-\omega_0)-i\mathcal{P}\frac{1}{\omega-\omega_0}\bigg]\\
&\quad+\sum_{j}\tilde{G}_{\alpha j}^*\sigma_{j}^\dagger (t)e^{i(\omega+\omega_0)t}
\bigg[\pi\delta(\omega+\omega_0)-i\mathcal{P}\frac{1}{\omega+\omega_0}\bigg],
\end{split}
\end{equation}
where the notation $\mathcal{P}$ stands for the principal value.
We insert Eq.~\eqref{11a} into Eq. (\ref{12}) and neglect fast-oscillating terms. 
Then, we obtain the Heisenberg-Langevin equation 
\begin{equation}\label{13}
\begin{split}
\dot{\mathcal{O}}&=\frac{1}{2}\sum_{ij}\gamma_{ij}(
[\sigma_{i}^{\dagger},\mathcal{O}] \sigma_{j}
+\sigma_{i}^{\dagger}[\mathcal{O},\sigma_{j}])\\
&+i\sum_{ij}\Omega_{ij}[\sigma_{i}^\dagger\sigma_{j},\mathcal{O}]
+\xi_{\mathcal{O}}.
\end{split}
\end{equation}
Here, the coefficients
\begin{equation}\label{14}
\gamma_{ij}=2\pi\sum_{\alpha}G_{\alpha i}G_{\alpha j}^*\delta(\omega-\omega_0)
\end{equation}
and
\begin{equation}\label{14a}
\Omega_{ij}=-\mathcal{P}\sum_{\alpha}\left[\frac{G_{\alpha i}G_{\alpha j}^*}{\omega-\omega_0}
+(-1)^{\delta_{ij}}\frac{\tilde{G}_{\alpha i}^*\tilde{G}_{\alpha j}}{\omega+\omega_0}\right]
\end{equation}
describe the decay rates and frequency shifts, respectively,
and $\xi_{\mathcal{O}}$ is the noise operator.

Let $\rho$ be  the reduced density operator for the atomic system.
When we use the Heisenberg-Langevin equation (\ref{13}) and the relation 
$\mathrm{Tr}[\mathcal{O}(t)\rho(0)]=\mathrm{Tr}[\mathcal{O}(0)\rho(t)]$, we find the master equation
\begin{eqnarray}\label{d21}
\dot{\rho}&=&\frac{1}{2}\sum_{ij}\gamma_{ij}(
2\sigma_{j}\rho\sigma_{i}^{\dagger}
-\sigma_{i}^{\dagger}\sigma_{j}\rho-\rho\sigma_{i}^{\dagger}\sigma_{j})\nonumber\\
&&\mbox{}-i\sum_{ij}\Omega_{ij}[\sigma_{i}^\dagger\sigma_{j},\rho].
\end{eqnarray}
In deriving the above equation, we multiplied Eq. (\ref{13}) with $\rho(0)$, took the trace of the result,
replaced the form $\mathrm{Tr}[\mathcal{O}_1(t)\mathcal{O}(t)\mathcal{O}_2(t)\rho(0)]$ by the form $\mathrm{Tr}[\mathcal{O}_1(0)\mathcal{O}(0)\mathcal{O}_2(0)\rho(t)]$, transformed to move the operator $\mathcal{O}(0)$ to the first position in each operator product, and eliminated $\mathcal{O}(0)$.

Note that $\gamma_{ij}=\gamma_{ji}^{*}$ and $\Omega_{ij}=\Omega_{ji}^{*}$. The single-atom coefficients $\gamma_{jj}$ and $\Omega_{jj}$ are real parameters. However, the cross-atom decay coefficient $\gamma_{12}$ and the dipole-dipole interaction coefficient $\Omega_{12}$ are generally complex parameters in the case of arbitrarily polarized dipoles. 

For two identical atoms with linearly polarized dipoles in free space, the cross-atom decay coefficient $\gamma_{12}$ and the dipole-dipole interaction coefficient $\Omega_{12}$ are real. In this case, the populations of the superradiant and subradiant superposition states decay with the rates
$\gamma_0+|\gamma_{12}|$ and $\gamma_0-|\gamma_{12}|$, respectively \cite{Agarwal book}. Here, $\gamma_0$ is the rate of single-atom decay
in free space. Meanwhile, the energy splitting between the superradiant  and subradiant states
is determined by the dipole-dipole coupling coefficient $\Omega_{12}$ \cite{Agarwal book}. 

The above interpretation remains valid when the cross-atom decay coefficient $\gamma_{12}$ and the dipole-dipole interaction coefficient $\Omega_{12}$ are complex parameters but have the same phase. Indeed, we can perform an appropriate transformation for the atomic operators to remove the phases of $\gamma_{12}$ and $\Omega_{12}$ if these phases are equal to each other. 

When the cross-atom decay coefficient $\gamma_{12}$ and the dipole-dipole interaction coefficient $\Omega_{12}$ are complex parameters and have different phases, it is not easy to interpret the physical meaning of these coefficients individually. 
Indeed, the imaginary part of the complex cross-atom decay coefficient $\gamma_{12}$ may affect the energy splitting between the superradiant and subradiant states, while the imaginary part of the complex dipole-dipole interaction coefficient $\Omega_{12}$ may affect the collective decay of atomic population. 

The roles of the absolute value and phase of the cross-atom decay coefficient $\gamma_{12}$ can be seen when we neglect the dipole-dipole interaction coefficient $\Omega_{12}$. In this case, the phase of $\gamma_{12}$ determines the relative phases between the component states $|+-\rangle$
and $|-+\rangle$ in the superradiant (symmetric) and subradiant (antisymmetric) superposition states, which are defined as the eigenstates of the collective atomic decay operator. Meanwhile, the absolute value of $\gamma_{12}$ determines the modifications of the decay rates of the superradiant and subradiant states, caused by the collective effect. 

The roles of the absolute value and phase of the dipole-dipole interaction coefficient $\Omega_{12}$ can be seen when we neglect the cross-atom decay coefficient $\gamma_{12}$. In this case, the phase of $\Omega_{12}$ determines the relative phases between the component states $|+-\rangle$
and $|-+\rangle$ in the one-excitation dressed states, which are defined as the eigenstates of the dipole-dipole interaction operator. Meanwhile, the absolute value of $\Omega_{12}$ determines the energy splitting between these dressed states.

In order to get deeper insight into the roles of the absolute values and phases of the complex collective coupling coefficients $\gamma_{12}$ and $\Omega_{12}$, we perform the following analysis:

Let $\gamma_{12}=|\gamma_{12}|\exp(i\phi_{\gamma})$ and $\Omega_{12}=|\Omega_{12}|\exp(i\phi_{\Omega})$, where
$\phi_{\gamma}$ and $\phi_{\Omega}$ are the phases of the complex coefficients $\gamma_{12}$ and $\Omega_{12}$, respectively.
We introduce the transformations $\tilde{\sigma}_1=\sigma_1\exp(i\phi_0)$ and  $\tilde{\sigma}_2=\sigma_2\exp[i(\phi_{\gamma}+\phi_0)]$,
where $\phi_0$ is an arbitrary parameter. Then, we can rewrite Eq.~\eqref{d21} as
\begin{eqnarray}\label{d21a}
\dot{\rho}&=&\frac{1}{2}\sum_{ij}\tilde{\gamma}_{ij}(
2\tilde{\sigma}_{j}\rho\tilde{\sigma}_{i}^{\dagger}
-\tilde{\sigma}_{i}^{\dagger}\tilde{\sigma}_{j}\rho-\rho\tilde{\sigma}_{i}^{\dagger}\tilde{\sigma}_{j})\nonumber\\
&&\mbox{}-i\sum_{ij}\tilde{\Omega}_{ij}[\tilde{\sigma}_{i}^\dagger\tilde{\sigma}_{j},\rho],
\end{eqnarray}
where $\tilde{\gamma}_{jj}=\gamma_{jj}$, $\tilde{\Omega}_{jj}=\Omega_{jj}$, $\tilde{\gamma}_{12}=|\gamma_{12}|$,
and $\tilde{\Omega}_{12}=|\Omega_{12}|\exp[i(\phi_{\Omega}-\phi_{\gamma})]$. 
It is clear that $\tilde{\gamma}_{jj}$, $\tilde{\Omega}_{jj}$, and $\tilde{\gamma}_{12}$ are real. However, 
when $|\Omega_{12}|\not=0$ and $\phi_{\Omega}-\phi_{\gamma}\not=0,\pm\pi$, the imaginary part of 
$\tilde{\Omega}_{12}$ is nonzero. It can be shown that the expression on the right-hand side of  
Eq.~\eqref{d21a} contains the different direction-dependent excitation transfer terms  $\tilde{\sigma}_1^{\dagger}\tilde{\sigma}_2\rho$ and $\tilde{\sigma}_2^{\dagger}\tilde{\sigma}_1\rho$ with the different coefficients $\tilde{\gamma}_{12}-2\mathrm{Im}(\tilde{\Omega}_{12})$ and $\tilde{\gamma}_{12}+2\mathrm{Im}(\tilde{\Omega}_{12})$, respectively.
When $\tilde{\gamma}_{12}\not=0$ and $\mathrm{Im}(\tilde{\Omega}_{12})\not=0$, the left-right symmetry is broken. Thus, when $|\gamma_{12}|\not=0$,
$|\Omega_{12}|\not=0$, and $\phi_{\Omega}-\phi_{\gamma}\not=0,\pm\pi$, the interaction between the atoms through the field depends on the direction of energy transfer, i.e., it is chiral \cite{Stannigel2012,Ramos2014,Zoller2015,Gonzalez-Ballestero2015,Ramos2016,Vermersch2016}.
We note that, in the particular case where $\phi_{\Omega}-\phi_{\gamma}=\pi/2$ and $|\Omega_{12}|=|\gamma_{12}|/2$, we have 
$\mathrm{Im}(\tilde{\Omega}_{12})=\tilde{\gamma}_{12}/2$. In this case, 
the expression on the right-hand side of  
Eq.~\eqref{d21a} contains the forward (left-to-right) excitation transfer term  $\tilde{\sigma}_2^{\dagger}\tilde{\sigma}_1\rho$
but not the backward (right-to-left) excitation transfer term $\tilde{\sigma}_1^{\dagger}\tilde{\sigma}_2\rho$ \cite{Stannigel2012,Ramos2014,Zoller2015,Gonzalez-Ballestero2015,Ramos2016,Vermersch2016}.  

We can write
\begin{equation}\label{15}
\begin{split}
\gamma_{ij}&=\gamma_{ij}^{(\mathrm{g})}+\gamma_{ij}^{(\mathrm{r})},\\
\Omega_{ij}&=\Omega_{ij}^{(\mathrm{g})}+\Omega_{ij}^{(\mathrm{r})},
\end{split}
\end{equation}
where the pair of $\gamma_{ij}^{(\mathrm{g})}$ and $\Omega_{ij}^{(\mathrm{g})}$
and the pair of $\gamma_{ij}^{(\mathrm{r})}$ and $\Omega_{ij}^{(\mathrm{r})}$ 
describe the contributions from guided and radiation modes, respectively. 
The coefficients  $\gamma_{ij}^{(\mathrm{g})}$ and $\gamma_{ij}^{(\mathrm{r})}$  
are given by
\begin{equation}\label{16}
\begin{split}
\gamma_{ij}^{(\mathrm{g})}&=2\pi \sum_{fl}G_{\mu_0 i}G_{\mu_0 j}^*,\\
\gamma_{ij}^{(\mathrm{r})}&=2\pi \int\limits _{-k_0}^{k_0}d\beta\sum_{ml}G_{\nu_0 i}G_{\nu_0 j}^*,
\end{split}
\end{equation}
where $\mu_0=(\omega_0,f,l)$ and $\nu_0=(\omega_0,\beta,m,l)$
label the resonant guided and radiation modes, whose frequencies coincide with the atomic frequency $\omega_0$. 
The coefficients  $\Omega_{ij}^{(\mathrm{g})}$ and $\Omega_{ij}^{(\mathrm{r})}$  
are given by
\begin{eqnarray}\label{16a}
\Omega_{ij}^{(\mathrm{g})}&=& 
-\mathcal{P}\int\limits _0^{\infty}d\omega\sum_{fl}
\left[\frac{G_{\omega fl i}G_{\omega fl j}^*}{\omega-\omega_0}+(-1)^{\delta_{ij}}
\frac{\tilde{G}_{\omega fl i}^*\tilde{G}_{\omega fl j}}{\omega+\omega_0}
\right],\nonumber\\
\Omega_{ij}^{(\mathrm{r})}&=& 
-\mathcal{P}\int\limits _0^{\infty}d\omega\sum_{ml}\int\limits _{-k}^{k}d\beta\,
\bigg[\frac{G_{\omega\beta ml i}G_{\omega\beta ml j}^*}{\omega-\omega_0}\nonumber\\
&&\mbox{} +(-1)^{\delta_{ij}}\frac{\tilde{G}_{\omega\beta ml i}^*\tilde{G}_{\omega\beta ml j}}{\omega+\omega_0}\bigg].
\end{eqnarray}

The directional components $\gamma_{ij}^{(\mathrm{g})f}$ of the rate $\gamma_{ij}^{(\mathrm{g})}$ for guided modes are given as
\begin{equation}\label{91a}
\gamma_{ij}^{(\mathrm{g})f}=2\pi\sum_{l} G_{\omega_0 fl i}G_{\omega_0 fl j}^*.
\end{equation}
The directional components $\gamma_{ij}^{(\mathrm{r})\pm}$ of the rate $\gamma_{ij}^{(\mathrm{r})}$ for radiation modes are given as
\begin{equation}\label{91b}
\begin{split}
\gamma_{ij}^{(\mathrm{r})+}&=2\pi \int\limits _{0}^{k_0}d\beta\sum_{ml}G_{\nu_0 i}G_{\nu_0 j}^*,\\
\gamma_{ij}^{(\mathrm{r})-}&=2\pi \int\limits _{-k_0}^{0}d\beta\sum_{ml}G_{\nu_0 i}G_{\nu_0 j}^*.
\end{split} 
\end{equation}

We note that, when the atoms are in free space, the decay rates and the dipole-dipole interaction coefficients are given as \cite{Lehmberg70,Agarwal92}
\begin{eqnarray}\label{99}
\gamma_{ij}^{(\mathrm{vac})}&=&\frac{\omega_0^3}{2\pi\hbar\epsilon_0c^3}\bigg\{
[\mathbf{d}_i\mathbf{d}_j^*-3(\mathbf{d}_i\cdot\hat{\mathbf{R}}_{ij})(\mathbf{d}_j^*\cdot\hat{\mathbf{R}}_{ij})]\nonumber\\
&&\mbox{}
\times\bigg(\frac{\cos k_0R_{ij}}{k_0^2R_{ij}^2}-\frac{\sin k_0R_{ij}}{k_0^3R_{ij}^3} \bigg)\nonumber\\
&&\mbox{}
+[\mathbf{d}_i\mathbf{d}_j^*-(\mathbf{d}_i\cdot\hat{\mathbf{R}}_{ij})(\mathbf{d}_j^*\cdot\hat{\mathbf{R}}_{ij})]
\frac{\sin k_0R_{ij}}{k_0R_{ij}}\bigg\}\qquad
\end{eqnarray}
and
\begin{equation}\label{100}
\begin{split}
\Omega_{ij}^{(\mathrm{vac})}\big|_{i\not=j}&=
\frac{\omega_0^3}{4\pi\hbar\epsilon_0c^3}\bigg\{
[\mathbf{d}_i\mathbf{d}_j^*-3(\mathbf{d}_i\cdot\hat{\mathbf{R}}_{ij})(\mathbf{d}_j^*\cdot\hat{\mathbf{R}}_{ij})]\\
&\quad
\times\bigg(\frac{\sin k_0R_{ij}}{k_0^2R_{ij}^2}+\frac{\cos k_0R_{ij}}{k_0^3R_{ij}^3} \bigg)\\
&\quad
-[\mathbf{d}_i\mathbf{d}_j^*-(\mathbf{d}_i\cdot\hat{\mathbf{R}}_{ij})(\mathbf{d}_j^*\cdot\hat{\mathbf{R}}_{ij})]
\frac{\cos k_0R_{ij}}{k_0R_{ij}}\bigg\}.
\end{split}
\end{equation}
Here, we have introduced the notation $\hat{\mathbf{R}}_{ij}=\mathbf{R}_{ij}/R_{ij}$ and $R_{ij}=|\mathbf{R}_{ij}|$, where $\mathbf{R}_{ij}=\mathbf{R}_i-\mathbf{R}_j$. According to Eqs.~\eqref{99} and \eqref{100}, the single-atom free-space coefficients $\gamma_{ii}^{(\mathrm{vac})}$ and $\Omega_{ii}^{(\mathrm{vac})}$ are real. It is clear from Eqs.~\eqref{99} and \eqref{100} that, when the two atoms have the same dipole matrix element, that is, when $\mathbf{d}_1=\mathbf{d}_2\equiv\mathbf{d}$, the cross-atom free-space coefficients $\gamma_{12}^{(\mathrm{vac})}$ and $\Omega_{12}^{(\mathrm{vac})}$ are also real. Thus,
the interaction between the atoms with the identical dipole matrix element in free space is not chiral.

\subsection{Dipole-dipole interaction}

As already mentioned above, the coefficients $\Omega_{ij}$ describe the frequency shifts of the two-atom system.
The diagonal coefficients $\Omega_{jj}$ describe the shifts of individual atoms. These shifts contain the Lamb shift and the surface-induced potential.
The Lamb shift can be formally incorporated into the bare frequency $\omega_0$. When the atoms are not very close to the surface, the surface-induced potential
is small. We are not interested in the surface-induced potential in this paper. Therefore, we neglect the diagonal coefficients $\Omega_{jj}$.
The off-diagonal coefficients $\Omega_{ij}=\Omega_{ji}^*$, where $i\not=j$, describe the dipole-dipole interaction between the atoms.

We calculate the coefficient $\Omega_{12}=\Omega_{21}^*$.
According to Eqs.~\eqref{16a}, we have
\begin{eqnarray}\label{16b}
\Omega_{12}^{(\mathrm{g})}&=& 
-\mathcal{P}\int\limits _0^{\infty}d\omega\sum_{fl}
\left(\frac{G_{\omega fl 1}G_{\omega fl 2}^*}{\omega-\omega_0}+\frac{\tilde{G}_{\omega fl 1}^*\tilde{G}_{\omega fl 2}}{\omega+\omega_0}
\right),\nonumber\\
\Omega_{12}^{(\mathrm{r})}&=& 
-\mathcal{P}\int\limits _0^{\infty}d\omega\sum_{ml}\int\limits _{-k}^{k}d\beta
\bigg(\frac{G_{\omega\beta ml 1}G_{\omega\beta ml 2}^*}{\omega-\omega_0}\nonumber\\
&&\mbox{} +\frac{\tilde{G}_{\omega\beta ml 1}^*\tilde{G}_{\omega\beta ml 2}}{\omega+\omega_0}\bigg).
\end{eqnarray}
We formally extend the field frequency $\omega$ from the region $[0,\infty]$ to the region $[-\infty,\infty]$. For guided modes, we 
use the definitions $\beta(-\omega)=-\beta(\omega)$ and $\mathbf{e}^{(-\omega,f,-l)}=\mathbf{e}^{(\omega,f,l)*}$. 
For radiation modes, we use the definition $\mathbf{e}^{(-\omega,-\beta,-m,l)}=\mathbf{e}^{(\omega,\beta,m,l)*}$.
These definitions are consistent with the time reversal symmetry of the Maxwell equations. 
With the aforementioned definitions, we have $G_{-\omega,f,-l,1}G_{-\omega,f,-l,2}^*=-\tilde{G}_{\omega fl1}^*\tilde{G}_{\omega fl2}$
and $G_{-\omega,-\beta,-m,l,1}G_{-\omega,-\beta,-m,l,2}^*=-\tilde{G}_{\omega\beta ml 1}^*\tilde{G}_{\omega\beta ml 2}$.
Then, Eqs.~\eqref{16b} become
\begin{subequations}\label{16c}
\begin{align}
\Omega_{12}^{(\mathrm{g})}&=-\mathcal{P}\int\limits_{-\infty}^{\infty}d\omega\sum_{fl}
\frac{G_{\omega fl 1}G_{\omega fl 2}^*}{\omega-\omega_0},\label{16ca}\\
\Omega_{12}^{(\mathrm{r})}&=-\mathcal{P}\int\limits_{-\infty}^{\infty}d\omega\sum_{ml}\int\limits _{-k}^{k}d\beta
\frac{G_{\omega\beta ml 1}G_{\omega\beta ml 2}^*}{\omega-\omega_0}.\label{16cb}
\end{align}
\end{subequations}

In the case of the waveguide bath models considered in Refs.~\cite{Stannigel2012,Ramos2014,Zoller2015,Gonzalez-Ballestero2015}, the radiation modes are not taken into account,
a single polarization guided modes is considered, and the coupling coefficient $G_{\omega fl j}$ for guided modes is 
replaced by $\sqrt{\gamma_f/2\pi}e^{if\omega z_j/v_g}$. Here, $\gamma_f$ is the decay rate into the direction $f$ of the waveguide axis. 
In this case, the dipole-dipole interaction coefficient is found from Eq.~\eqref{16ca} to be \cite{Stannigel2012,Ramos2014,Zoller2015,Gonzalez-Ballestero2015}
\begin{equation}\label{16c1}
\Omega_{12}^{(1D)}\big|_{z_1\not=z_2}=-\frac{1}{2\pi}\sum_{f}\gamma_f\;\mathcal{P}\int\limits_{-\infty}^{\infty}d\omega
\frac{e^{if\omega z_{12}/v_g}}{\omega-\omega_0}.
\end{equation}
Here, $z_{ij}=z_i-z_j$ is the difference between the axial positions of atoms $i$ and $j$. 
When we use the contour integral method to calculate the integral over $\omega$ in Eq.~\eqref{16c1}, we obtain  \cite{Stannigel2012,Ramos2014,Zoller2015,Gonzalez-Ballestero2015}
\begin{equation}\label{16c2}
\Omega_{12}^{(1D)}\big|_{z_1\not=z_2}=-\frac{i}{2} \sum_{f}\mathrm{sign}(fz_{12})\gamma_{f}e^{if\omega_0 z_{12}/v_g}.
\end{equation}

In the case of nanofibers, we can use the contour integral method to calculate approximately the integral over $\omega$ in Eq.~\eqref{16ca} for $\Omega_{12}^{(\mathrm{g})}$. For this purpose, we need to choose an appropriate close contour consisting of the line segments $(-R,\omega_0-\epsilon)$ and $(\omega_0+\epsilon,R)$ and two semicircles $C_\epsilon$ and $C_R$ connecting the point $\omega_0-\epsilon$ with the point $\omega_0+\epsilon$  and the point $R$ with the point $-R$, respectively. Here, $\epsilon>0$ is a small real number and $R>0$ is a large real number.
The semicircle $C_R$ lies in the upper or lower half plane of $\omega$ depending on the asymptotic behavior of the integral kernel 
$G_{\omega fl 1}G_{\omega fl 2}^*$. According to Eq.~\eqref{8}, the product $G_{\omega fl 1}G_{\omega fl 2}^*$ contains the factor $e^{if\beta(z_1-z_2)}$. We assume that $z_1\not=z_2$ and that the $\omega$ dependence of $G_{\omega fl 1}G_{\omega fl 2}^*$ is mainly determined by the factor $e^{if\beta(z_1-z_2)}$. With an appropriate choice of the half plane to place $C_R$, we can make the integral over this semicircle vanishing. The integral over the small semicircle $C_\epsilon$ can be calculated by using the residue theorem.
Then, we find
\begin{equation}\label{16d}
\Omega_{12}^{(\mathrm{g})}\big|_{z_1\not=z_2}\simeq 
-\pi i\sum_{fl}\mathrm{sign}(fz_{12}) G_{\omega_0 fl 1}G_{\omega_0 fl 2}^*.
\end{equation}
We can rewrite Eq.~\eqref{16d} in the form  \cite{Stannigel2012,Ramos2014,Zoller2015,Gonzalez-Ballestero2015}
\begin{equation}\label{d91}
\Omega_{12}^{(\mathrm{g})}\big|_{z_1\not=z_2}\simeq  -\frac{i}{2} \sum_{f}\mathrm{sign}(fz_{12})\gamma_{12}^{(\mathrm{g})f},
\end{equation}
where $\gamma_{12}^{(\mathrm{g})f}$ is the cross-atom decay coefficient for the $f$ propagation direction.
It is clear that Eq.~\eqref{d91} is in agreement with Eq.~\eqref{16c2}.
We can formally extend Eq.~\eqref{d91} for the case of $z_1=z_2$ by taking the limit $z_2\to z_1$ under the condition $z_2>z_1$.

We note that it is not easy to calculate the integral over $\omega$ in Eq.~\eqref{16cb} for $\Omega_{12}^{(\mathrm{r})}$. 
The reason is that the $\omega$ dependence of the integral kernel $G_{\omega\beta ml 1}G_{\omega\beta ml 2}^*$ is complicated.

\subsection{Photon flux}

The mean number of photons in guided modes propagating the direction $f=\pm$ is given by
\begin{equation}\label{d73} 
N_{\mathrm{gyd}}^{(f)}=\sum_l\int_0^{\infty} \langle a_{\omega fl}^\dagger a_{\omega fl}\rangle d\omega.
\end{equation}
The mean number of emitted guided-mode photons, summed up over the propagation directions, is $N_{\mathrm{gyd}}=N_{\mathrm{gyd}}^{(+)}+N_{\mathrm{gyd}}^{(-)}$.
The flux of photons emitted into the guided modes in the direction $f=\pm$ is given by 
\begin{equation}\label{d74} 
P_{\mathrm{gyd}}^{(f)}=\dot{N}_{\mathrm{gyd}}^{(f)}=\sum_l\int_0^{\infty} \langle \dot{a}_{\omega fl}^\dagger a_{\omega fl}+a_{\omega fl}^\dagger \dot{a}_{\omega fl}\rangle d\omega. 
\end{equation}

We insert Eqs.~\eqref{10} and \eqref{11a} into Eq.~\eqref{d74} and neglect the fast rotating terms. Then, we obtain
\begin{equation}\label{d75}
P_{\mathrm{gyd}}^{(f)}=\sum_{ij}\gamma_{ij}^{(\mathrm{g})f}\langle\sigma^{\dagger}_{i}\sigma_{j}\rangle,
\end{equation}
that is,
\begin{eqnarray}\label{d76}
P_{\mathrm{gyd}}^{(f)}&=&\gamma_{11}^{(\mathrm{g})f}\langle\sigma^{\dagger}_{1}\sigma_{1}\rangle+\gamma_{22}^{(\mathrm{g})f}\langle\sigma^{\dagger}_{2}\sigma_{2}\rangle 
+ \gamma_{12}^{(\mathrm{g})f}\langle\sigma^{\dagger}_{1}\sigma_{2}\rangle\nonumber\\
&&\mbox{} +\gamma_{21}^{(\mathrm{g})f}\langle\sigma^{\dagger}_{2}\sigma_{1}\rangle.
\end{eqnarray}
In terms of the density matrix $\rho$, Eq.~\eqref{d76} can be rewritten as
\begin{equation}\label{d77}
\begin{split}
P_{\mathrm{gyd}}^{(f)}&=\gamma_{11}^{(\mathrm{g})f}(\rho_{++,++}+\rho_{+-,+-})\\
&\quad +\gamma_{22}^{(\mathrm{g})f}(\rho_{++,++}+\rho_{-+,-+})\\
&\quad +\gamma_{12}^{(\mathrm{g})f}\rho_{-+,+-}+
\gamma_{21}^{(\mathrm{g})f}\rho_{+-,-+}.
\end{split}
\end{equation}

The flux $P_{\mathrm{gyd}}=P_{\mathrm{gyd}}^{(+)}+P_{\mathrm{gyd}}^{(-)}$ of photons emitted into guided modes in the two directions $f=\pm$ is given as
\begin{eqnarray}\label{d77a}
P_{\mathrm{gyd}}&=&\gamma_{11}^{(\mathrm{g})}(\rho_{++,++}+\rho_{+-,+-})+\gamma_{22}^{(\mathrm{g})}(\rho_{++,++}+\rho_{-+,-+})\nonumber\\
&&\mbox{}+\gamma_{12}^{(\mathrm{g})}\rho_{-+,+-}+
\gamma_{21}^{(\mathrm{g})}\rho_{+-,-+}.
\end{eqnarray}
Similarly, the flux of photons emitted into radiation modes is given by
\begin{eqnarray}\label{d77b}
P_{\mathrm{rad}}&=&\gamma_{11}^{(\mathrm{r})}(\rho_{++,++}+\rho_{+-,+-})+\gamma_{22}^{(\mathrm{r})}(\rho_{++,++}+\rho_{-+,-+})\nonumber\\
&&\mbox{}+\gamma_{12}^{(\mathrm{r})}\rho_{-+,+-}+
\gamma_{21}^{(\mathrm{r})}\rho_{+-,-+}.
\end{eqnarray}
The mean number of photons emitted into radiation modes is $N_{\mathrm{rad}}(t)=\int_{t_0}^{t}P_{\mathrm{rad}}(t')dt'$.

The total flux $P_{\mathrm{tot}}=P_{\mathrm{gyd}}+P_{\mathrm{rad}}$ of photons emitted into guided and radiation modes is given as
\begin{eqnarray}\label{d77c}
P_{\mathrm{tot}}&=&\gamma_{11}(\rho_{++,++}+\rho_{+-,+-})+\gamma_{22}(\rho_{++,++}+\rho_{-+,-+})\nonumber\\
&&\mbox{}+\gamma_{12}\rho_{-+,+-}+
\gamma_{21}\rho_{+-,-+}.
\end{eqnarray}
The mean number of photons emitted into guided and radiation modes is $N_{\mathrm{tot}}(t)=\int_{t_0}^{t}P_{\mathrm{tot}}(t')dt'$.
It can be shown that 
\begin{equation}\label{d77d}
P_{\mathrm{tot}}=-\dot{\rho}_{\mathrm{exc}},
\end{equation}
 where $\rho_{\mathrm{exc}}=\rho_{\mathrm{exc}}^{(1)}+\rho_{\mathrm{exc}}^{(2)}$
with $\rho_{\mathrm{exc}}^{(j)}=\langle \sigma_j^\dagger\sigma_j \rangle$ being the population of  the excited level of  atom $j$.

It is clear that the coefficients of the terms in the expressions for the photon fluxes are the single- and cross-atom decay coefficients. The dipole-dipole interaction coefficients do not enter these expressions explicitly.

\section{Numerical calculations}
\label{sec:numerical}

In what follows, we present the results of our numerical calculations pertaining to the decay rates,  
the dipole-dipole interaction coefficients, the time dependences of the populations of the atomic excited states, and the fluxes and mean numbers
of emitted guided-mode photons. Since the case of real dipole matrix elements has been studied \cite{twoatoms}, 
we consider here the case where the dipole matrix elements are complex vectors. In this case, spontaneous emission and scattering of light may become asymmetric with respect to the opposite axial propagation directions \cite{Fam14,Petersen14,Mitsch14b,AtomArray,Scheel15,Sayrin15b}.
The directionality of emission from a single atom occurs when the atomic dipole matrix element vector
is a complex vector in the plane that contains the fiber axis and the atomic position \cite{Fam14}. 
To be specific, we assume that the atomic transitions are $\sigma_+$-polarized transitions with respect to the $y$ axis,
that is, the dipole matrix elements of the atoms are $\mathbf{d}_j=(d/\sqrt2)(i,0,-1)$ for $j=1,2$.
In our numerical calculations, we take the fiber radius 
$a=250$ nm and the wavelength of the atomic transition $\lambda_0=852$ nm. 
The refractive indices of the fiber and the surrounding vacuum are $n_1=1.45$ and 
$n_2=1$, respectively. The single- and cross-atom decay coefficients will be compared to the decay rate 
$\gamma_0=\omega_0^3d^2/(3\pi\hbar\epsilon_0 c^3)$ of a single atom in free space.

\subsection{Decay rates}
\label{subsection:rates}

\begin{figure}[tbh]
\begin{center}
  \includegraphics{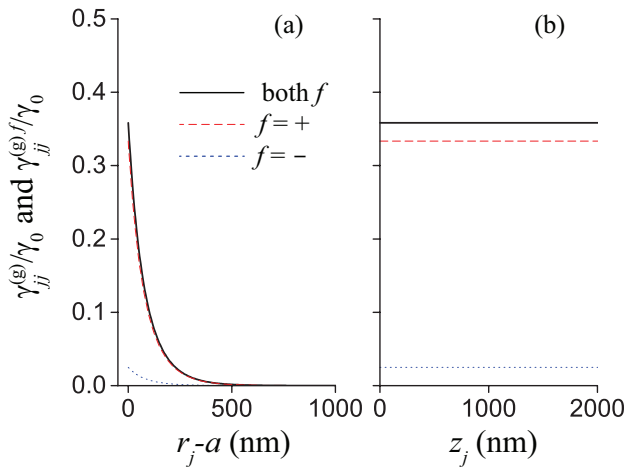}
 \end{center}
\caption{
Decay rate $\gamma_{jj}^{(\mathrm{g})}$ into guided modes (solid black lines) and its directional components
$\gamma_{jj}^{(\mathrm{g})f}$ for $f=+$ (dashed red lines) and $f=-$ (dotted blue lines), relative to the free-space spontaneous decay rate $\gamma_0$, as functions of (a) the radial position $r_j-a$ and (b) the axial position $z_j$ of atom $j$. One coordinate of the atom is varied, while the two others are fixed as (a) $\varphi_j=0$ and $z_j=0$ and (b) $r_j=a$ and $\varphi_j=0$. The dipole matrix element of the atom is $\mathbf{d}_j=(d/\sqrt2)(i,0,-1)$, corresponding to the $\sigma_+$-polarized transition with respect to the $y$ quantization axis. The fiber radius is $a=250$ nm. The refractive indices of the fiber and the surrounding vacuum are $n_1=1.45$ and 
$n_2=1$, respectively. The wavelength of the atomic transition is $\lambda_0=852$ nm.
}
\label{fig2}
\end{figure}

\begin{figure}[tbh]
\begin{center}
  \includegraphics{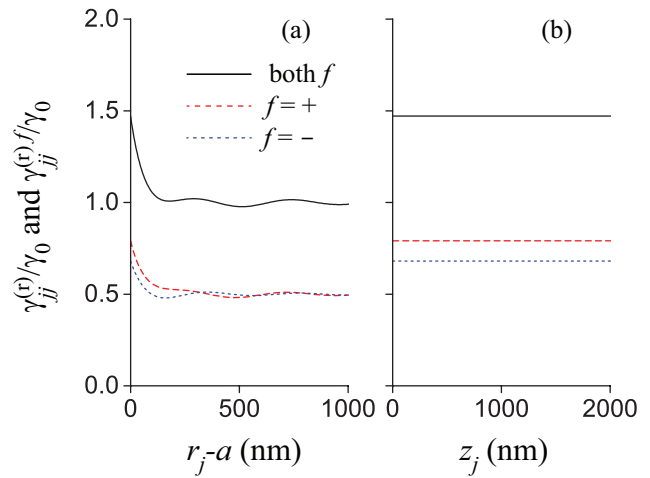}
 \end{center}
\caption{
Decay rate $\gamma_{jj}^{(\mathrm{r})}$ into radiation modes (solid black lines) and its directional components
$\gamma_{jj}^{(\mathrm{r})f}$ for $f=+$ (dashed red lines) and $f=-$ (dotted blue lines), relative to the free-space spontaneous decay rate $\gamma_0$, as functions of (a) the radial position $r_j-a$ and (b) the axial position $z_j$ of atom $j$. Other parameters are as for Fig.~\ref{fig2}.
}
\label{fig3}
\end{figure}

\begin{figure}[tbh]
\begin{center}
  \includegraphics{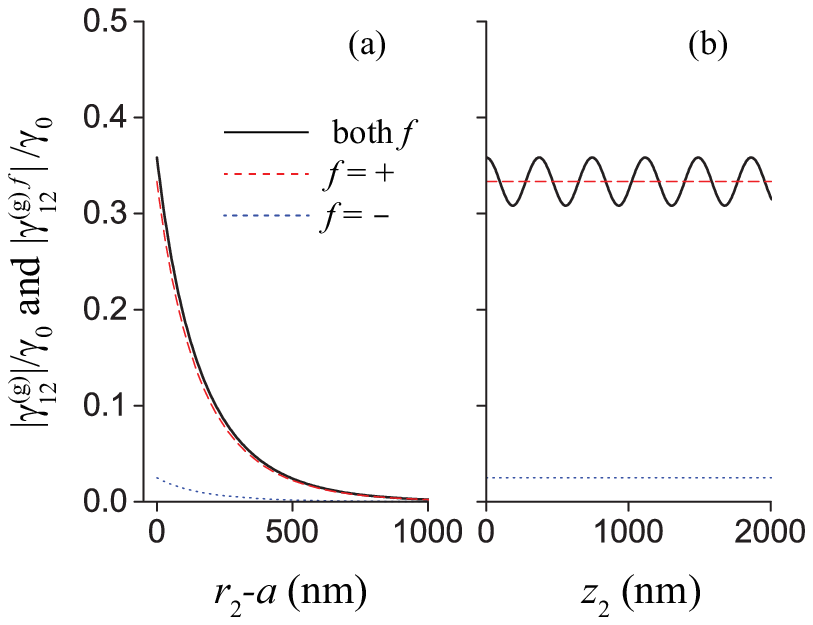}
 \end{center}
\caption{
Absolute value $|\gamma_{12}^{(\mathrm{g})}|$ of the coefficient of the cross-atom decay into guided modes (solid black lines) and its directional components
$|\gamma_{12}^{(\mathrm{g})f}|$ for $f=+$ (dashed red lines) and $f=-$ (dotted blue lines), relative to the free-space spontaneous decay rate $\gamma_0$, as functions of (a) the radial position $r_2-a$ and (b) the axial position $z_2$ of atom $2$. 
The position of atom $1$ is fixed at $r_1=a$, $\varphi_1=0$, and $z_1=0$.
One coordinate of atom $2$ is varied, while the two others are fixed as (a) $\varphi_2=0$ and $z_2=0$ and (b) $r_2=a$ and $\varphi_2=0$. The dipole matrix elements of both atoms are $\mathbf{d}_1=\mathbf{d}_2=(d/\sqrt2)(i,0,-1)$, corresponding to the $\sigma_+$-polarized transitions with respect to the $y$ quantization axis. Other parameters are as for Fig.~\ref{fig2}.
}
\label{fig4}
\end{figure}

\begin{figure}[tbh]
\begin{center}
  \includegraphics{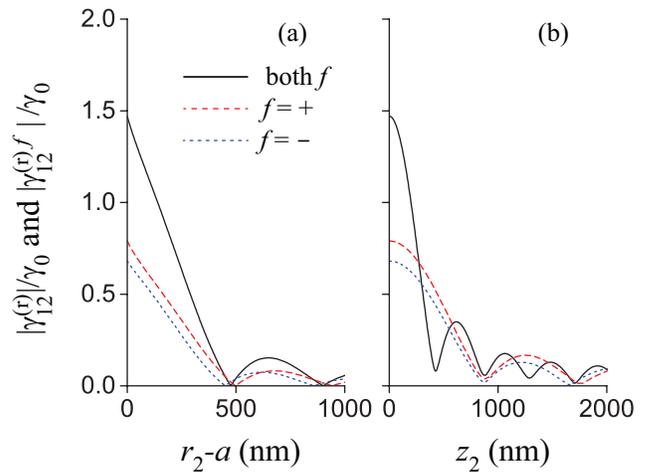}
 \end{center}
\caption{
Absolute value $|\gamma_{12}^{(\mathrm{r})}|$ of the coefficient of the cross-atom decay into radiation modes (solid black lines) and its directional components
$|\gamma_{12}^{(\mathrm{r})f}|$ for $f=+$ (dashed red lines) and $f=-$ (dotted blue lines), relative to the free-space spontaneous decay rate $\gamma_0$, as functions of (a) the radial position $r_2-a$ and (b) the axial position $z_2$ of atom $2$. Other parameters are as for Figs.~\ref{fig2} and \ref{fig4}.
}
\label{fig5}
\end{figure}

We calculate the single-atom decay rates $\gamma_{jj}^{(\mathrm{g})}$ and $\gamma_{jj}^{(\mathrm{r})}$
into guided and radiation modes, respectively, and the cross-atom decay coefficients $\gamma_{12}^{(\mathrm{g})}$ and $\gamma_{12}^{(\mathrm{r})}$
into guided and radiation modes, respectively, as functions of the radial and axial positions of the atoms. 
We plot the single-atom decay rates $\gamma_{jj}^{(\mathrm{g})}$ and $\gamma_{jj}^{(\mathrm{r})}$ in Figs. \ref{fig2} and \ref{fig3}, respectively.  
We plot the absolute values $|\gamma_{12}^{(\mathrm{g})}|$ and $|\gamma_{12}^{(\mathrm{r})}|$ of the cross-atom decay coefficients $\gamma_{12}^{(\mathrm{g})}$ and $\gamma_{12}^{(\mathrm{r})}$, respectively, in Figs. \ref{fig4} and \ref{fig5}, respectively. We also plot the directional components $\gamma_{jj}^{(\mathrm{g})f}$ and $\gamma_{jj}^{(\mathrm{r})f}$ of the rates $\gamma_{jj}^{(\mathrm{g})}$ and $\gamma_{jj}^{(\mathrm{r})}$, respectively,
in Figs. \ref{fig2} and \ref{fig3}, respectively, and the absolute values of  the directional components $\gamma_{12}^{(\mathrm{g})f}$ and $\gamma_{12}^{(\mathrm{r})f}$ of the rates $\gamma_{12}^{(\mathrm{g})}$ and $\gamma_{12}^{(\mathrm{r})}$, respectively, in Figs. \ref{fig4} and \ref{fig5}, respectively.
Parts (a) and (b) of Figs.~\ref{fig2}--\ref{fig5} stand for the dependences of the rates on the radial and axial positions of the atoms, respectively. The dotted blue, dashed red, and solid black curves refer to the rates for the negative ($f=-$) direction, the positive ($f=+$) direction, and the sum of the  rates for the two opposite directions, respectively. Comparison between the dashed red and dotted blue curves shows that the rates are different for the
opposite axial directions $f=+$ and $f=-$. The asymmetry is due to the existence of a nonzero longitudinal component of the nanofiber field, which is in phase quadrature with respect to the radial transverse component  \cite{Fam14,Petersen14,Mitsch14b,AtomArray,Scheel15,Sayrin15b}. 
This asymmetry occurs when the ellipticity vector of the atomic dipole polarization overlaps with the ellipticity vector of the field polarization \cite{Fam14,Fam16}. The directional spontaneous emission is a signature of spin-orbit coupling of light carrying transverse spin angular momentum \cite{Zeldovich,Bliokh review,Bliokh review2015,Bliokh2014,Bliokh2015,Banzer review2015}. We observe from Fig.~\ref{fig2}(a) that,
for the parameters of the figure, we have $\gamma_{jj}^{(\mathrm{g})+}>\gamma_{jj}^{(\mathrm{g})-}$, that is, spontaneous emission into guided modes in the positive direction $f=+$ is stronger than that in the negative direction $f=-$. The dominance of spontaneous emission into guided modes in the direction $f=+$ occurs for any radial distance $r$ in the case considered. Meanwhile, Fig.~\ref{fig3}(a) shows that, in spontaneous emission into radiation modes, both the possibilities $\gamma_{jj}^{(\mathrm{r})+}>\gamma_{jj}^{(\mathrm{r})-}$ and $\gamma_{jj}^{(\mathrm{r})+}<\gamma_{jj}^{(\mathrm{r})-}$ may appear, depending on the radial distance $r$ \cite{Scheel15}. The dependences of the rates $\gamma_{jj}^{(\mathrm{r})f}$ and  $\gamma_{12}^{(\mathrm{r})f}$ for radiation modes (see Figs.~\ref{fig3} and \ref{fig5}) on the emission direction $f$ are, in general, weaker than those of the rates $\gamma_{jj}^{(\mathrm{g})f}$ and $\gamma_{12}^{(\mathrm{g})f}$ for guided modes (see Figs.~\ref{fig2} and \ref{fig4}), respectively. 
 
The results presented in Figs.~\ref{fig2} and \ref{fig3} are in perfect agreement with the results of Refs.~\cite{Fam14,Scheel15}. The steep reductions of the decay rates $\gamma_{jj}^{(\mathrm{g})}$ and $\gamma_{12}^{(\mathrm{g})}$ with increasing radial distance in Figs.~\ref{fig2}(a) and \ref{fig4}(a), respectively, are the consequences of the evanescent-wave nature of the field in the guided modes. The single-atom decay rates $\gamma_{jj}^{(\mathrm{g})}$ and $\gamma_{jj}^{(\mathrm{r})}$  do not depend on the axial position $z_j$ [see Figs.~\ref{fig2}(b) and \ref{fig3}(b)]. Meanwhile, the cross-atom decay coefficients $\gamma_{12}^{(\mathrm{g})}$ and $\gamma_{12}^{(\mathrm{r})}$ oscillate with increasing axial separation between the atoms  [see Figs.~\ref{fig4}(b) and \ref{fig5}(b)]. It can be easily discerned from Figs.~\ref{fig4}(b) and \ref{fig5}(b) that the effect of guided modes on the cross-atom decay persists over arbitrarily large axial separations between the atoms while that due to the radiation modes decays to zero. 
Thus the guided modes of the fiber play a crucial role in maintaining the coupling over large distances \cite{twoatoms}. 
It is clear that one can control the coupling between the atoms by varying the separation between them with maximum coupling at certain locations. 
We observe from Fig.~\ref{fig4}(b) that the cross-atom guided-mode-mediated decay coefficient $\gamma_{12}^{(\mathrm{g})}$ oscillates with increasing axial separation but does not cross the zero value axis. This behavior is a consequence of chiral coupling between the atoms.
Indeed, in the case considered, we have $|\gamma_{12}^{(\mathrm{g})+}|>|\gamma_{12}^{(\mathrm{g})-}|$. 
Meanwhile, $\gamma_{12}^{(\mathrm{g})+}$ and $\gamma_{12}^{(\mathrm{g})-}$ are complex parameters, whose dependences on the axial coordinates of the atoms are given by the factors $\exp(i\beta_0 z_{12})$ and $\exp(-i\beta_0 z_{12})$, respectively.
Since $|\gamma_{12}^{(\mathrm{g})+}|>|\gamma_{12}^{(\mathrm{g})-}|$, the interference between $\gamma_{12}^{(\mathrm{g})+}$ and $\gamma_{12}^{(\mathrm{g})-}$ can never be completely destructive. Thus, in the case of chiral coupling, the cross-atom guided-mode-mediated decay coefficient $\gamma_{12}^{(\mathrm{g})}$ is nonzero for arbitrary values $z_{12}$. It is worth noting that in the case of nonchiral coupling \cite{twoatoms},
$\gamma_{12}^{(\mathrm{g})}$ vanishes when $\beta_0z=\pi/2+n\pi$, where $n=0,1,2,\dots$.

\begin{figure}[tbh]
\begin{center}
  \includegraphics{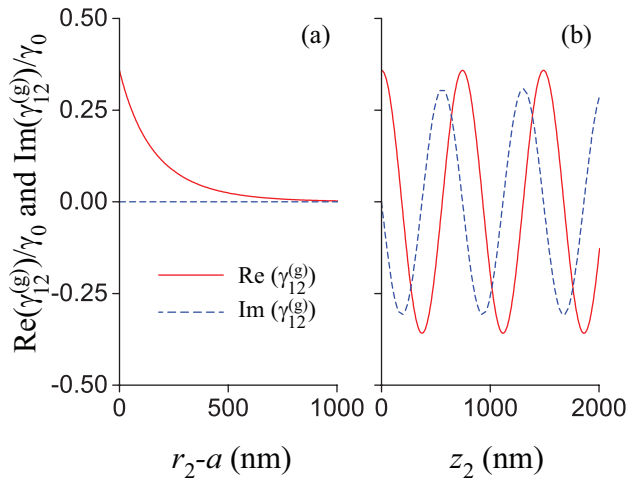}
 \end{center}
\caption{
Real (solid red lines) and imaginary (dashed blue lines) parts of the coefficient $\gamma_{12}^{(\mathrm{g})}$ for cross-atom decay into guided modes, relative to the free-space spontaneous decay rate $\gamma_0$, as functions of (a) the radial position $r_2-a$ and (b) the axial position $z_2$ of atom $2$. 
Other parameters are as for Figs.~\ref{fig2} and \ref{fig4}.
}
\label{fig6}
\end{figure}

\begin{figure}[tbh]
\begin{center}
  \includegraphics{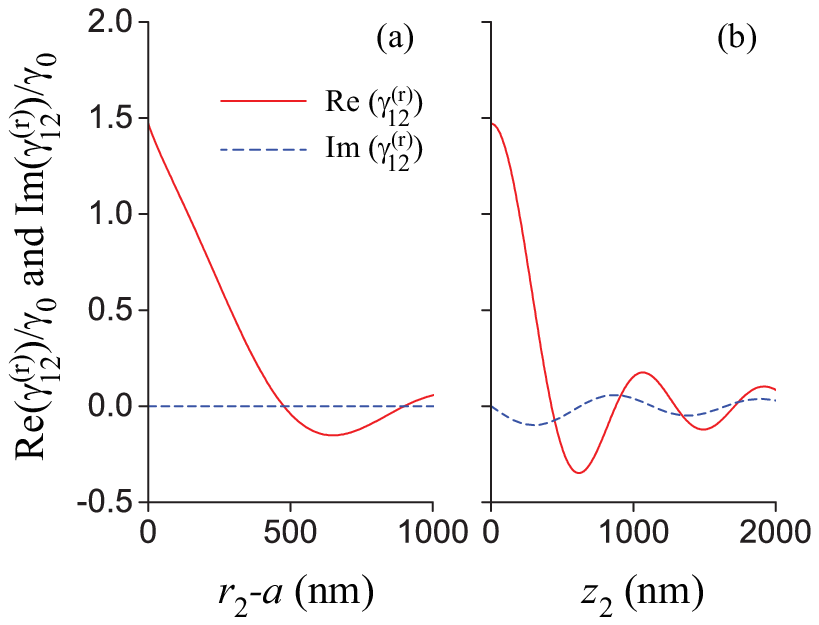}
 \end{center}
\caption{
Real (solid red lines) and imaginary (dashed blue lines) parts of the coefficient $\gamma_{12}^{(\mathrm{r})}$ for cross-atom decay into radiation modes, relative to the free-space spontaneous decay rate $\gamma_0$, as functions of (a) the radial position $r_2-a$ and (b) the axial position $z_2$ of atom $2$. 
Other parameters are as for Figs.~\ref{fig2} and \ref{fig4}.
}
\label{fig7}
\end{figure}

\begin{figure}[tbh]
\begin{center}
  \includegraphics{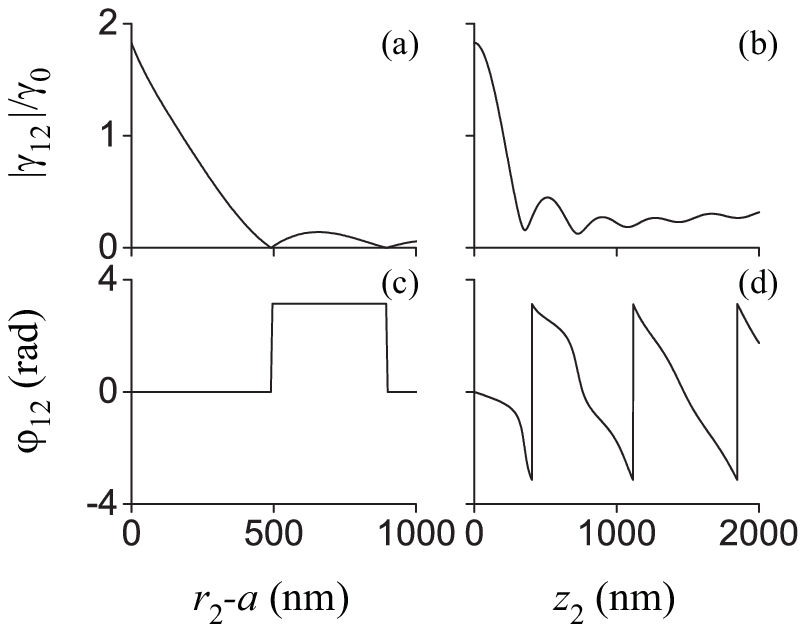}
 \end{center}
\caption{
Absolute value $|\gamma_{12}|$ (upper row) and phase $\varphi_{12}$ (lower row) of the total cross-atom decay coefficient $\gamma_{12}$
as functions of the radial position $r_2-a$ (left column) and the axial position $z_2$ (right column) of atom $2$. 
Other parameters are as for Figs.~\ref{fig2} and \ref{fig4}.
}
\label{fig8}
\end{figure}

We note that, in the case where the dipole matrix elements $\mathbf{d}_j$ are complex vectors, the cross-atom decay coefficients $\gamma_{12}^{(\mathrm{g})}$ and $\gamma_{12}^{(\mathrm{r})}$ into guided and radiation modes, respectively, are, in general, complex parameters. In order to illustrate this feature, we plot separately the real and imaginary parts of $\gamma_{12}^{(\mathrm{g})}$ in Fig.~\ref{fig6} and the real and imaginary parts of $\gamma_{12}^{(\mathrm{r})}$ in Fig.~\ref{fig7}. In addition, we plot in Fig.~\ref{fig8} the absolute value $|\gamma_{12}|$ and the phase $\varphi_{12}$ of the total cross-atom decay coefficient $\gamma_{12}=\gamma_{12}^{(\mathrm{g})}+\gamma_{12}^{(\mathrm{r})}$. Figures \ref{fig6}(a) and \ref{fig6}(b) show respectively the evanescent-wave behavior of the radial dependence and the oscillatory behavior of the axial dependence of the cross-atom coefficient $\gamma_{12}^{(\mathrm{g})}$ of decay into guided modes. We observe from Fig.~\ref{fig6}(b) that the real and imaginary parts of $\gamma_{12}^{(\mathrm{g})}$ oscillate periodically with a relative phase difference of $\pi/2$ along the fiber axis. Figure \ref{fig7} shows that the cross-atom coefficient $\gamma_{12}^{(\mathrm{r})}$ of decay into radiation modes oscillates in the radial and axial directions and that the amplitude of oscillations reduces with increasing separation between the atoms. Figure \ref{fig8}(a) indicates the possibility of the channels of decay into guided and radiation modes to act out of phase, leading to $\gamma_{12}=0$ at certain points. We observe from Figs.~\ref{fig8}(c) and \ref{fig8}(d) that the phase $\varphi_{12}$ of the total cross-atom decay coefficient $\gamma_{12}$ depends on the positions of the atoms.

\subsection{Dipole-dipole interaction coefficients}
\label{subsection:dipole-dipole}

\begin{figure}[tbh]
\begin{center}
  \includegraphics{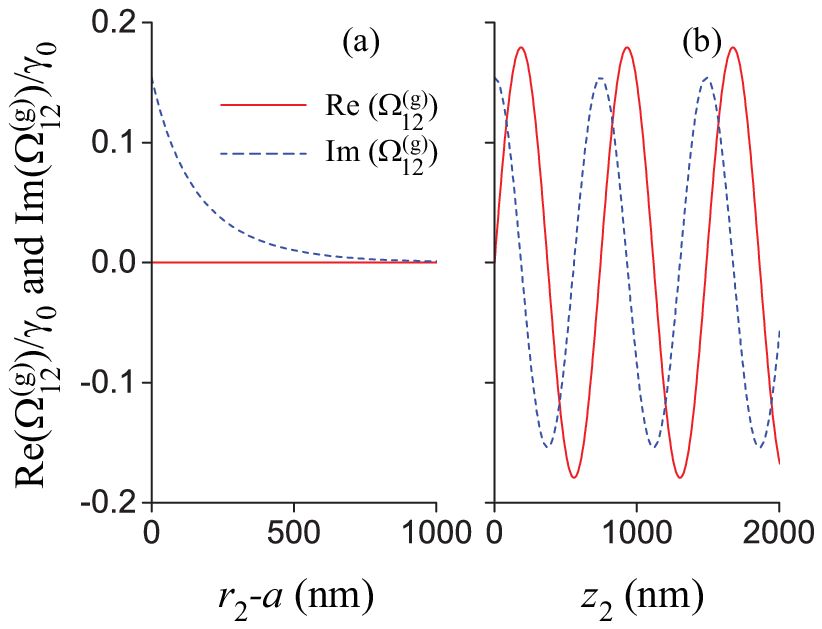}
 \end{center}
\caption{
Real (solid red lines) and imaginary (dashed blue lines) parts of the guided-mode-mediated dipole-dipole interaction coefficient $\Omega_{12}^{(\mathrm{g})}$, relative to the free-space spontaneous decay rate $\gamma_0$, as functions of (a) the radial position $r_2-a$ and (b) the axial position $z_2$ of atom $2$. Parameters used are as for Figs.~\ref{fig2} and \ref{fig4}.
In part (a), we formally take the limit $z_2\to z_1$ under the condition $z_2>z_1$.
}
\label{fig9}
\end{figure}

We plot in Fig.~\ref{fig9} the real and imaginary parts of the guided-mode-mediated dipole-dipole interaction coefficient $\Omega_{12}^{(\mathrm{g})}$.
Figures \ref{fig9}(a) and \ref{fig9}(b) show respectively the evanescent-wave behavior of the radial dependence and the oscillatory behavior of the axial dependence of the coefficient $\Omega_{12}^{(\mathrm{g})}$. We observe from Fig.~\ref{fig9}(b) that the real and imaginary parts of 
$\Omega_{12}^{(\mathrm{g})}$ oscillate periodically with a relative phase difference of $\pi/2$ along the fiber axis. 
  
\begin{figure}[tbh]
\begin{center}
  \includegraphics{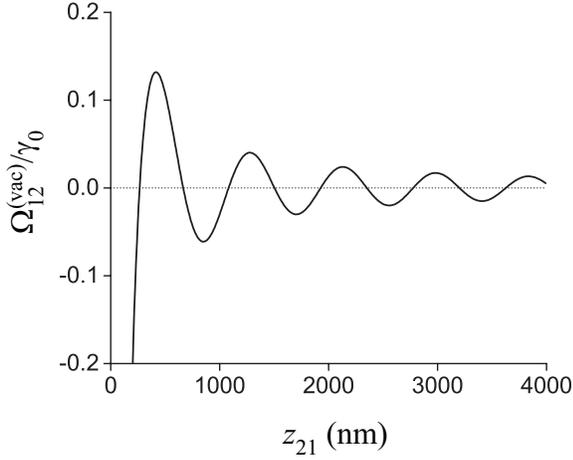}
 \end{center}
\caption{
Free-space dipole-dipole interaction coefficient $\Omega_{12}^{(\mathrm{vac})}$, relative to the free-space spontaneous decay rate $\gamma_0$, as a function of the distance $z_{21}=z_2-z_1$ between the atoms along the fiber axis. 
The other coordinates of the atoms  are $r_1=r_2$ and $\varphi_1=\varphi_2$.
The dipole matrix elements of the atoms are $\mathbf{d}_1=\mathbf{d}_2=(d/\sqrt2)(i,0,-1)$, corresponding to the $\sigma_+$-polarized transitions with respect to the $y$ quantization axis. The dotted line is the zero horizontal axis and is a guide to the eye.}
\label{fig10}
\end{figure} 

\begin{figure}[tbh]
\begin{center}
  \includegraphics{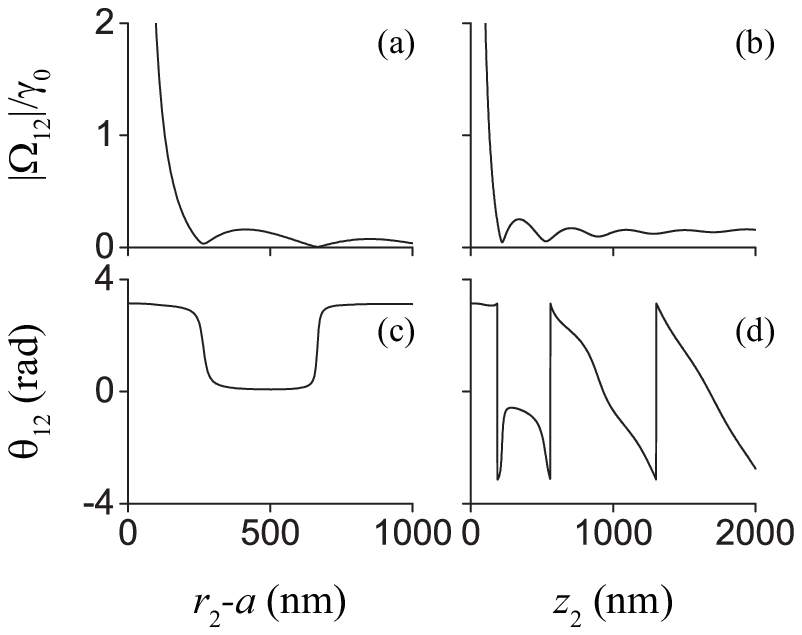}
 \end{center}
\caption{
Absolute value $|\Omega_{12}|$ (upper row) and phase $\theta_{12}$ (lower row) of the total dipole-dipole interaction coefficient 
$\Omega_{12}$ as functions of the radial position $r_2-a$ (left column) and the axial position $z_2$ (right column) of atom $2$. 
Other parameters are as for Figs.~\ref{fig2} and \ref{fig4}.}
\label{fig11}
\end{figure} 

The expression \eqref{16cb} for the radiation-mode-mediated dipole-dipole interaction coefficient $\Omega_{12}^{(\mathrm{r})}$ contains a double integral
and a double sum of Bessel functions. It is not easy to calculate numerically this coefficient. 
When the atoms are not too close to the fiber surface, the effect of the fiber on $\Omega_{12}^{(\mathrm{r})}$ is not serious. In this case,
$\Omega_{12}^{(\mathrm{r})}$ is close to $\Omega_{12}^{(\mathrm{vac})}$, where  $\Omega_{12}^{(\mathrm{vac})}$ is the dipole-dipole interaction coefficient
for atoms in free-space. We use the approximation $\Omega_{12}^{(\mathrm{r})}\simeq \Omega_{12}^{(\mathrm{vac})}\sqrt{\gamma_{11}^{(\mathrm{r})}\gamma_{22}^{(\mathrm{r})}}/\gamma_0$.  Here, we have added the factor $\sqrt{\gamma_{11}^{(\mathrm{r})}\gamma_{22}^{(\mathrm{r})}}/\gamma_0$ to take into account the effect of the fiber on the mode density of radiation modes. 
As already mentioned in the previous section, the free-space dipole-dipole interaction coefficient $\Omega_{12}^{(\mathrm{vac})}$ is real
in the case where the two atoms have the same dipole matrix element  ($\mathbf{d}_1=\mathbf{d}_2\equiv\mathbf{d}$).
We plot in Fig.~\ref{fig10} the coefficient $\Omega_{12}^{(\mathrm{vac})}$ as a function of the distance between the atoms.
We depict in Fig.~\ref{fig11} the absolute value $|\Omega_{12}|$ and the phase $\theta_{12}$ of the total dipole-dipole interaction coefficient $\Omega_{12}=\Omega_{12}^{(\mathrm{g})}+\Omega_{12}^{(\mathrm{r})}$. Figure \ref{fig10} shows that the free-space dipole-dipole interaction coefficient $\Omega_{12}^{(\mathrm{vac})}$ oscillates and decays with increasing separation between the atoms.
Figure \ref{fig11}(a) indicates that $\Omega_{12}$ becomes close to zero at certain positions of the atoms along the radial direction. 
This feature is due to the existence of zeros of $\Omega_{12}^{(\mathrm{vac})}$ (see Fig.~\ref{fig10}) and the quick reduction of $\Omega_{12}^{(\mathrm{g})}$ with increasing distance of one of the atoms to the fiber surface. 
We observe from Figs.~\ref{fig11}(c) and \ref{fig11}(d) that the phase $\theta_{12}$ of the total dipole-dipole interaction coefficient $\Omega_{12}$ depends on the positions of the atoms. Comparison between Figs.~\ref{fig8}(c) and \ref{fig11}(c) and between Figs.~\ref{fig8}(d) and \ref{fig11}(d) shows that the phases $\varphi_{12}$ and $\theta_{12}$ of the coefficients $\gamma_{12}$ and $\Omega_{12}$ are, in general, different from each other.

\subsection{Dynamics}
\label{subsection:dynamics}

We solve the master equation \eqref{d21} for different initial states.
We use the solutions of this equation to calculate the populations $\rho_{\mathrm{exc}}^{(j)}=\langle\sigma_j^\dagger\sigma_j\rangle$
of the upper levels of atoms $j=1,2$, the fluxes $P_{\mathrm{gyd}}^{(f)}$ of photons emitted into guided modes in the direction $f=\pm$ along the fiber axis, and the mean number $N_{\mathrm{gyd}}^{(f)}$ of photons emitted into guided modes in the direction $f$. 
We also calculate the total flux $P_{\mathrm{gyd}}=\sum_f P_{\mathrm{gyd}}^{(f)}$ 
and the total mean number $N_{\mathrm{gyd}}=\sum_f N_{\mathrm{gyd}}^{(f)}$ of photons emitted into guided modes.
We study first the cases where an atom is initially excited and the other atom is initially in the ground state and then the cases where the two atoms are prepared in a symmetric or antisymmetric superposition state.

\subsubsection{An excited atom in the presence of a ground-state atom}

We first study the cases where an atom is initially excited and the other atom is initially not excited. In these cases, 
the initial state of the two-atom system is $|\psi(0)\rangle=|\psi_1\rangle$ or $|\psi_2\rangle$, where
$|\psi_1\rangle\equiv |+-\rangle$ and $|\psi_2\rangle\equiv |-+\rangle$. The direction of radiative transfer in the case of the initial state
$|\psi_1\rangle$ or $|\psi_2\rangle$ is from atom 1 to atom 2 or from atom 2 to atom 1, respectively. 

\begin{figure}[tbh]
\begin{center}
  \includegraphics{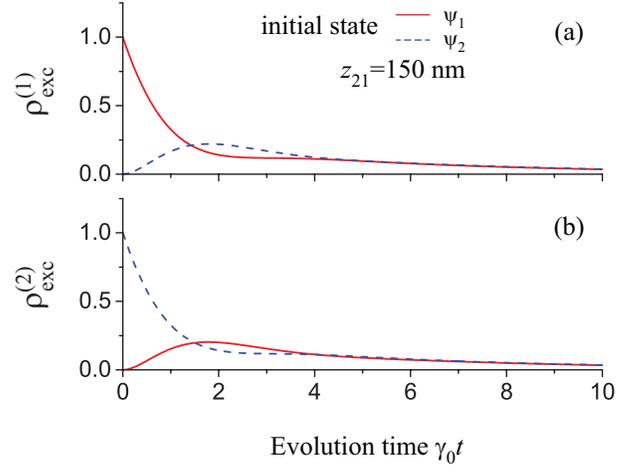}
 \end{center}
\caption{
Time evolution of the populations $\rho_{\mathrm{exc}}^{(1)}$ (a) and $\rho_{\mathrm{exc}}^{(2)}$ (b) of the upper levels of atoms 1 and 2, respectively,  in the cases where
the initial state of the two-atom system is $|\psi(0)\rangle=|\psi_{1}\rangle$ (solid red lines) or $|\psi_{2}\rangle$ (dashed blue lines).
The coordinates of the atoms are $r_1-a=r_2-a=200$ nm, $\varphi_1=\varphi_2=0$, and $z_2-z_1=150$ nm.
The dipole matrix elements of the atoms are $\mathbf{d}_1=\mathbf{d}_2=(d/\sqrt2)(i,0,-1)$, corresponding to $\sigma_+$-polarized transitions with respect to the $y$ quantization axis. Other parameters are as for Figs.~\ref{fig2} and \ref{fig4}.
}
\label{fig12}
\end{figure}

\begin{figure}[tbh]
\begin{center}
  \includegraphics{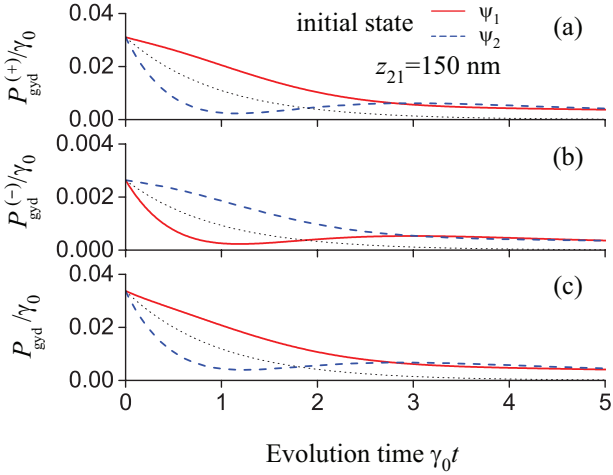}
 \end{center}
\caption{
Time evolution of the photon fluxes $P_{\mathrm{gyd}}^{(+)}$ (a), $P_{\mathrm{gyd}}^{(-)}$ (b), and $P_{\mathrm{gyd}}$ (c) in the cases where
the initial state of the two-atom system is $|\psi(0)\rangle=|\psi_{1}\rangle$ (solid red lines) or $|\psi_{2}\rangle$ (dashed blue lines).
Other parameters are as for Figs.~\ref{fig2}, \ref{fig4}, and \ref{fig12}.
The fluxes are normalized to the decay rate $\gamma_0$ of a single atom in free space. For comparison, the results for the case of a single excited atom are
shown by the dotted black lines.}
\label{fig13}
\end{figure}

\begin{figure}[tbh]
\begin{center}
  \includegraphics{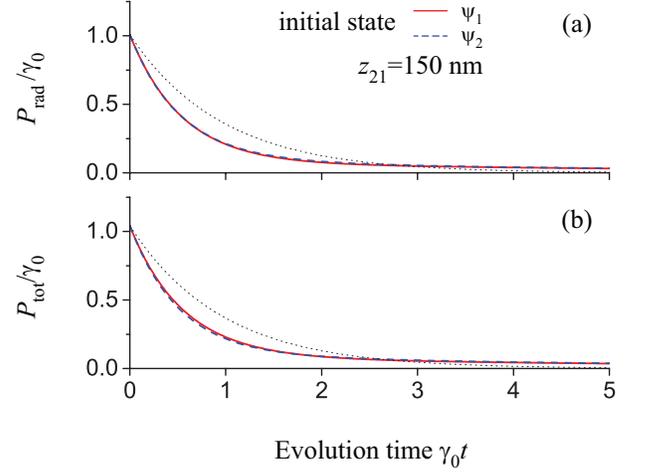}
 \end{center}
\caption{
Time evolution of the photon fluxes $P_{\mathrm{rad}}$ (a) and $P_{\mathrm{tot}}$ (b) in the cases where
the initial state of the two-atom system is $|\psi(0)\rangle=|\psi_{1}\rangle$ (solid red lines) or $|\psi_{2}\rangle$ (dashed blue lines).
Other parameters are as for Figs.~\ref{fig2}, \ref{fig4}, and \ref{fig12}. The fluxes are normalized to the decay rate $\gamma_0$ of a single atom in free space.
The dotted black lines are for the case of a single excited atom.}
\label{fig14}
\end{figure}

We plot in Figs.~\ref{fig12}--\ref{fig14} the results of numerical calculations for the case where the coordinates of the atoms are $r_1-a=r_2-a=200$ nm, $\varphi_1=\varphi_2=0$, and $z_2-z_1=150$ nm.

Figure \ref{fig12} shows the time evolution of the populations $\rho_{\mathrm{exc}}^{(j)}$ of
the excited states of the atoms in the cases where
the initial state of the two-atom system is $|\psi(0)\rangle=|\psi_1\rangle$ (solid red lines) or $|\psi_2\rangle$ (dashed blue lines).
We observe in both cases
that a part of the atomic excitation is transferred from the excited atom to the ground-state atom, and then is slowly released by emission. 
Comparison between the solid red and dashed blue lines of Fig.~\ref{fig12} shows that, except for the changes of the roles of the atoms, the differences between the results for the cases of the initial states $|\psi(0)\rangle=|\psi_1\rangle$ and $|\psi(0)\rangle=|\psi_2\rangle$ are very small.
Comparison between the solid red line of  Fig.~\ref{fig12}(a) and the dashed blue line of Fig.~\ref{fig12}(b) shows that the decay of $\rho_{\mathrm{exc}}^{(1)}$ in the case of $|\psi(0)\rangle=|\psi_1\rangle$ is almost the same as the decay of $\rho_{\mathrm{exc}}^{(2)}$ in the case of $|\psi(0)\rangle=|\psi_2\rangle$.
Meanwhile, a close inspection shows that the peak of the transferred excitation $\rho_{\mathrm{exc}}^{(2)}$ in Fig.~\ref{fig12}(b) (see the solid red line of this figure) is slightly different from the peak of the transferred excitation $\rho_{\mathrm{exc}}^{(1)}$ in Fig.~\ref{fig12}(a) (see the dashed blue line of this figure). Our additional calculations which are not shown here indicate that, depending on the parameters of the system, the peak of the transferred excitation $\rho_{\mathrm{exc}}^{(2)}$ in the case of the initial state $|\psi_1\rangle$ [see the solid red line of Fig.~\ref{fig12}(b)] may be slightly larger or smaller than the peak of the transferred excitation $\rho_{\mathrm{exc}}^{(1)}$ in the case of the initial state $|\psi_2\rangle$ 
[see the dashed blue line of Fig.~\ref{fig12}(a)]. 

Figure \ref{fig13} shows the time evolution of the fluxes $P_{\mathrm{gyd}}^{(+)}$ and $P_{\mathrm{gyd}}^{(-)}$ of photons emitted into guided modes in the positive and negative directions of the fiber axis, respectively, and the total guided-photon flux $P_{\mathrm{gyd}}$, calculated for the cases where the initial state of the two-atom system is $|\psi(0)\rangle=|\psi_{1}\rangle$ (solid red lines) or $|\psi_{2}\rangle$ (dashed blue lines). 
For comparison, the corresponding results for the case of a single excited atom are shown by the dotted black lines.
Comparison between the scales of the vertical axes in Figs.~\ref{fig13}(a) and \ref{fig13}(b) shows that the photon flux $P_{\mathrm{gyd}}^{(+)}$ for the positive direction is about one order larger than the photon flux $P_{\mathrm{gyd}}^{(-)}$ for the negative direction.
Furthermore, we observe that the photon fluxes for the initial states $|\psi_1\rangle$ (solid red lines) and $|\psi_2\rangle$ (dashed blue lines) are substantially different from each other.
Thus, the photon fluxes depend on the direction of propagation of light and the direction of radiative transfer between the atoms. 
We emphasize again that this is a chiral effect and is a signature of spin-orbit coupling of light \cite{Zeldovich,Bliokh review,Bliokh review2015,Bliokh2014,Bliokh2015,Banzer review2015}. This effect results from the existence of a nonzero longitudinal component of the nanofiber field, which is in phase quadrature with respect to the radial transverse component \cite{Fam14,Petersen14,Mitsch14b,AtomArray,Scheel15,Sayrin15b}. 

Comparison between the solid red, dashed blue, and  dotted  black lines of Fig.~\ref{fig13} shows that the presence of a ground-state atom in the vicinity
of an excited atom may increase or decrease the fluxes of photons emitted into guided modes. Thus, the collective emission into guided modes can be enhanced or
suppressed depending on the direction of propagation of light and the direction of radiative transfer between the atoms.
We note that the flux of emitted photons depends on not only the single-atom excited populations $\rho_{\mathrm{exc}}^{(1)}$ and $\rho_{\mathrm{exc}}^{(2)}$ but also on the cross-atom interference. In addition, the atoms can emit not only into guided modes but also into radiation modes.

The variation of the total atomic excitation $\rho_{\mathrm{exc}}=\rho_{\mathrm{exc}}^{(1)}+\rho_{\mathrm{exc}}^{(2)}$ in time is proportional to the total flux $P_{\mathrm{tot}}$ of photons emitted into guided and radiation modes [see Eq.~\eqref{d77d}]. We plot in Fig.~\ref{fig14} the time evolution of the flux $P_{\mathrm{rad}}$ of photons emitted into radiation modes and the total photon flux $P_{\mathrm{tot}}$ for the parameters of Fig.~\ref{fig13}. We observe that, unlike the flux $P_{\mathrm{gyd}}$ into guides modes, the flux $P_{\mathrm{rad}}$ into radiation modes and the total flux $P_{\mathrm{tot}}$ do not depend significantly on the direction of excitation transfer.
In addition, we observe that, when the interaction time is not zero and not too large, the fluxes $P_{\mathrm{rad}}$ and $P_{\mathrm{tot}}$ from two atoms in the initial state $|\psi(0)\rangle=|\psi_{1}\rangle$ (solid red lines) or $|\psi_{2}\rangle$ (dashed blue lines) are smaller than the corresponding fluxes from a single excited atom (dotted black lines). Such reductions of $P_{\mathrm{rad}}$ and $P_{\mathrm{tot}}$ are a consequence of the excitation transfer between the atoms. The effect of the cross-atom interference on the fluxes $P_{\mathrm{rad}}$ and $P_{\mathrm{tot}}$ is not as strong as that on the flux $P_{\mathrm{gyd}}$.

\begin{figure}[tbh]
\begin{center}
  \includegraphics{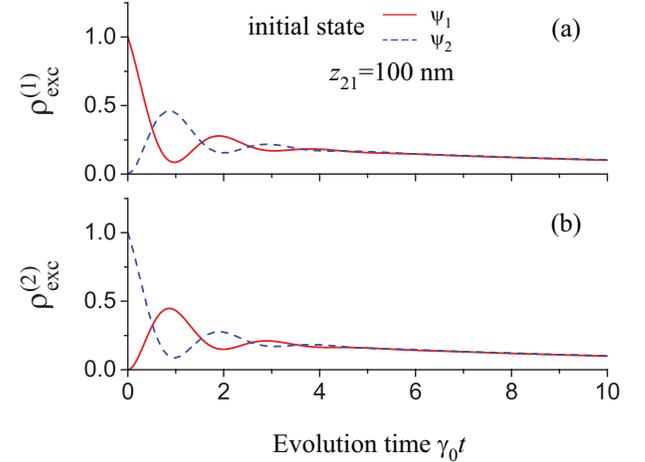}
 \end{center}
\caption{
Same as Fig.~\ref{fig12} except for $z_2-z_1=100$ nm.
}
\label{fig15}
\end{figure}

\begin{figure}[tbh]
\begin{center}
  \includegraphics{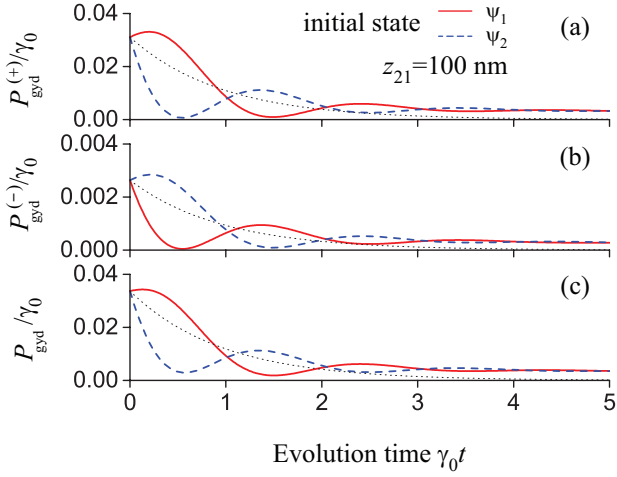}
 \end{center}
\caption{
Same as Fig.~\ref{fig13} except for $z_2-z_1=100$ nm.
}
\label{fig16}
\end{figure}

We note that, when we reduce the distance between the atoms, the dipole-dipole interaction increases. When this interaction is strong enough,  
we may observe oscillations in the time dependences of the excited-state populations $\rho_{\mathrm{exc}}^{(1)}$ and  $\rho_{\mathrm{exc}}^{(2)}$ and the photon fluxes $P_{\mathrm{gyd}}^{(+)}$, $P_{\mathrm{gyd}}^{(-)}$, and $P_{\mathrm{gyd}}$. In order to illustrate such a situation, we plot in Figs.~\ref{fig15} and \ref{fig16} the results of calculations
for the quantities presented in Figs.~\ref{fig12} and \ref{fig13}, respectively, using the same parameters except for $z_2-z_1=100$ nm.
We observe clearly oscillations in the time evolution of the calculated quantities.
For the parameters used, we do not see oscillations in $P_{\mathrm{rad}}$ and $P_{\mathrm{tot}}$.

\begin{figure}[tbh]
\begin{center}
  \includegraphics{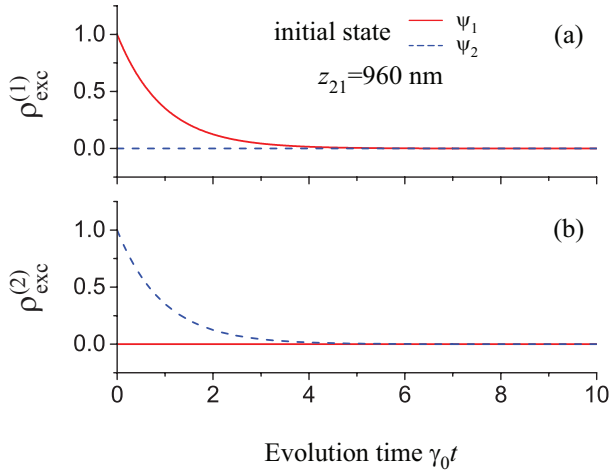}
 \end{center}
\caption{
Same as Fig.~\ref{fig12} except for $z_2-z_1=960$ nm.
}
\label{fig17}
\end{figure}

\begin{figure}[tbh]
\begin{center}
  \includegraphics{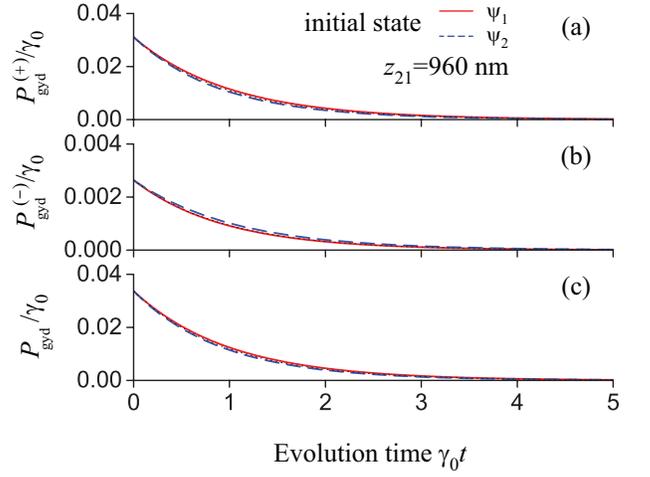}
 \end{center}
\caption{
Same as Fig.~\ref{fig13} except for $z_2-z_1=960$ nm.
}
\label{fig18}
\end{figure}

In the limit $z_{21}\to\infty$, the cross-atom radiation-mode-mediated coefficients $\gamma_{12}^{(\mathrm{r})}$ and $\Omega_{12}^{(\mathrm{r})}$ tend to vanish. In this limit, the collective effects are mainly determined  by the cross-atom guided-mode-mediated coefficients $\gamma_{12}^{(\mathrm{g})}$ and $\Omega_{12}^{(\mathrm{g})}$, which are, in general, finite. In order to illustrate such a situation, we plot in Figs.~\ref{fig17} and \ref{fig18} the results of calculations for the quantities presented in Figs.~\ref{fig12} and \ref{fig13}, respectively, using the same parameters except for $z_2-z_1=960$ nm. We observe from Fig.~\ref{fig17} that the transfer of excitation between the atoms is negligible. Figure \ref{fig18} shows that the differences between the results for the cases $|\psi(0)\rangle=|\psi_{1}\rangle$ (solid red lines) and $|\psi_{2}\rangle$ (dashed blue lines) are small but not negligible.

\begin{figure}[tbh]
\begin{center}
  \includegraphics{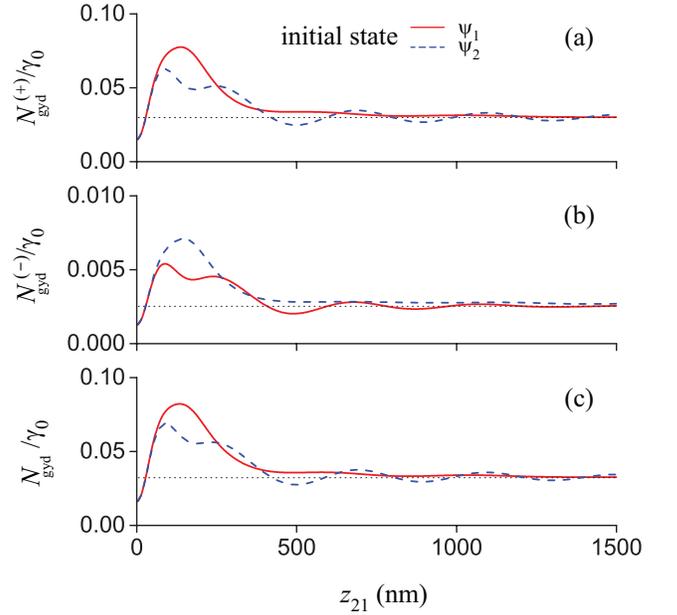}
 \end{center}
\caption{
Dependences of the mean emitted photon numbers $N_{\mathrm{gyd}}^{(+)}$ (a), $N_{\mathrm{gyd}}^{(-)}$ (b), and $N_{\mathrm{gyd}}$ (c) on the axial separation $z_{21}$ between the atoms in the cases where
the initial state of the two-atom system is $|\psi(0)\rangle=|\psi_{1}\rangle$ (solid red lines) or $|\psi_{2}\rangle$ (dashed blue lines). 
The radial and azimuthal coordinates of the atoms are $r_1-a=r_2-a=200$ nm and $\varphi_1=\varphi_2=0$, respectively.
Other parameters are as for Figs.~\ref{fig2}, \ref{fig4}, and \ref{fig12}.
The dotted black lines are for the case of a single excited atom.}
\label{fig19}
\end{figure}

\begin{figure}[tbh]
\begin{center}
  \includegraphics{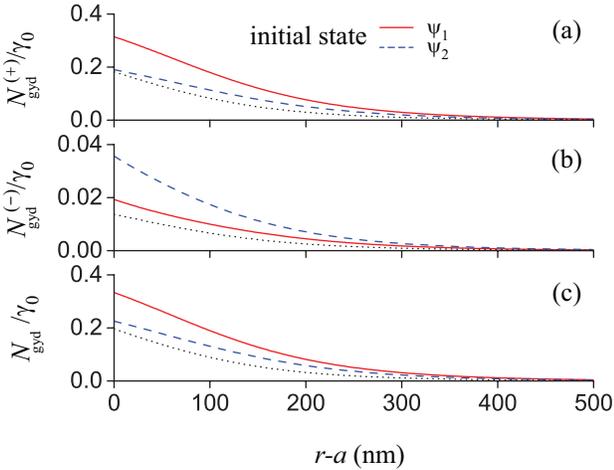}
 \end{center}
\caption{
Dependences of the mean emitted photon numbers $N_{\mathrm{gyd}}^{(+)}$ (a), $N_{\mathrm{gyd}}^{(-)}$ (b), and $N_{\mathrm{gyd}}$ (c) on the distance $r-a$ from the atoms to the fiber surface 
in the cases where the initial state of the two-atom system is $|\psi(0)\rangle=|\psi_{1}\rangle$ (solid red lines) or $|\psi_{2}\rangle$ (dashed blue lines). 
The coordinates of the atoms are $r_1=r_2=r$, $\varphi_1=\varphi_2=0$, and $z_2-z_1=150$ nm.
Other parameters are as for Figs.~\ref{fig2}, \ref{fig4}, and \ref{fig12}.
The dotted black lines are for the case of a single excited atom.}
\label{fig20}
\end{figure}

The mean numbers $N_{\mathrm{gyd}}^{(+)}$, $N_{\mathrm{gyd}}^{(-)}$, and $N_{\mathrm{gyd}}$ of photons emitted into guided modes in the positive direction, the negative direction, and both directions, respectively, are determined by the integrations of the fluxes  $P_{\mathrm{gyd}}^{(+)}$, $P_{\mathrm{gyd}}^{(-)}$, and $P_{\mathrm{gyd}}$, respectively, over the evolution time $t$. We plot in Figs.~\ref{fig19} and \ref{fig20} the dependences of the mean emitted guided photon numbers on the axial atomic separation $z_{21}$ and the atom-to-surface distance $r-a$, respectively.
The results for the cases of the initial states $|\psi(0)\rangle=|\psi_{1}\rangle$ and $|\psi_{2}\rangle$ are shown by the solid red lines and the dashed blue lines, respectively. For comparison, we plot the corresponding results for the case of a single excited atom by the dotted black lines.

Comparison between the scales of Figs.~\ref{fig19}(a) and \ref{fig19}(b) and between the scales of Figs.~\ref{fig20}(a) and \ref{fig20}(b) shows
that the mean photon number $N_{\mathrm{gyd}}^{(+)}$ for the positive direction is about one order larger than the mean photon number $N_{\mathrm{gyd}}^{(-)}$ for the negative direction.
It is clear from the figure that the mean emitted guided photon number $N_{\mathrm{gyd}}$ and its directional components $N_{\mathrm{gyd}}^{(+)}$ and $N_{\mathrm{gyd}}^{(-)}$ depend on the axial atomic separation $z_{21}$, the atom-surface distance $r-a$, and the direction of radiative transfer between the atoms. 

When we compare the solid red and dashed blue lines of Fig.~\ref{fig19} with the dotted black lines of this figure, we see that, depending on the axial atomic separation $z_{21}$ and the radiative transfer direction, the presence of a ground-state atom may enhance or suppress the probability for an excited atom to emit a photon into guided modes. In addition, we observe that, depending on $z_{21}$, the values of $N_{\mathrm{gyd}}^{(+)}$, $N_{\mathrm{gyd}}^{(-)}$, and $N_{\mathrm{gyd}}$ in the case of the initial state $|\psi(0)\rangle=|\psi_{1}\rangle$ (solid red lines) may be larger or smaller than the corresponding values in the case of the initial state $|\psi(0)\rangle=|\psi_{2}\rangle$ (dashed blue lines). For $z_{21}$ in the region from 25 to 400 nm, $N_{\mathrm{gyd}}$ and its directional components $N_{\mathrm{gyd}}^{(+)}$ and $N_{\mathrm{gyd}}^{(-)}$ for the two-atom case (see the solid red and dashed blue lines) are significantly larger than the corresponding values for a single excited atom (see the dotted black lines). These differences are signatures of the collective effect in spontaneous emission into guided modes. We note that an increase or a decrease in the mean number of photons emitted into guided modes is associated with a decrease or an increase, respectively, in the mean number of photons emitted into radiation modes.

\begin{figure}[tbh]
\begin{center}
  \includegraphics{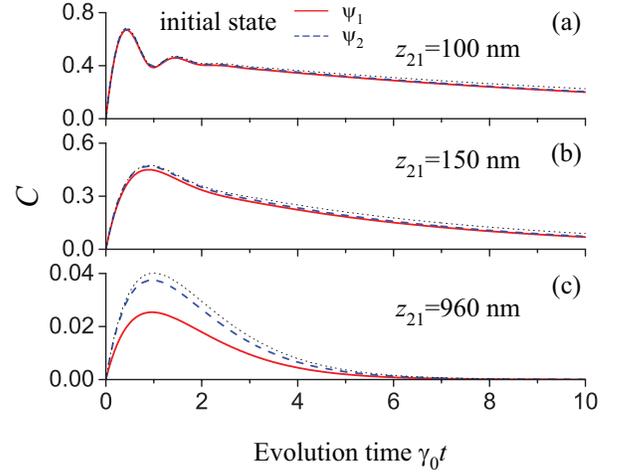}
 \end{center}
\caption{
Time evolution of the concurrence in the cases where
the initial state of the two-atom system is $|\psi(0)\rangle=|\psi_{1}\rangle$ (solid red lines) or $|\psi_{2}\rangle$ (dashed blue lines).
The coordinates of the atoms are $r_1-a=r_2-a=200$ nm, $\varphi_1=\varphi_2=0$, and $z_2-z_1=100$ nm (a), 150 nm (b), and 960 nm (c).
Other parameters are as for Figs.~\ref{fig2} and~ \ref{fig4}.
The dotted black lines are for the case of two atoms in free space.
}
\label{fig21}
\end{figure}

\begin{figure}[tbh]
\begin{center}
  \includegraphics{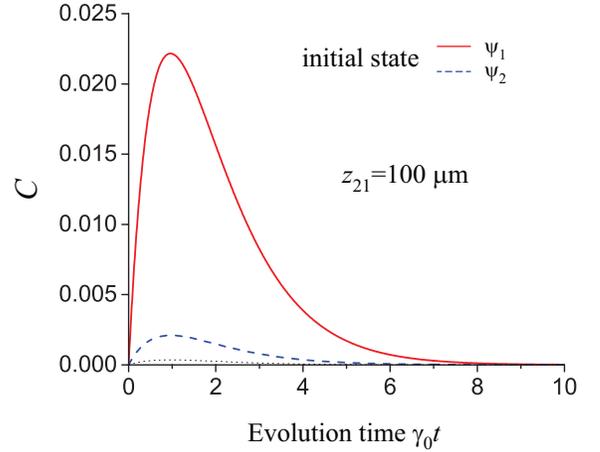}
 \end{center}
\caption{
Same as Fig.~\ref{fig21} except for $z_2-z_1=100$ $\mu$m.
}
\label{fig22}
\end{figure}

The interaction of the atoms prepared in the state $|\psi(0)\rangle=|\psi_{1}\rangle$ or $|\psi_{2}\rangle$ with the vacuum of the field may lead to entanglement between the atoms. The entanglement can be characterized by the concurrence $C$ \cite{Wootters}. For two two-level atoms, the density matrix elements are denoted as $\rho_{\alpha\beta}$, where $\alpha,\beta=e,g,a,b$ with $|e\rangle=|++\rangle$, $|g\rangle=|--\rangle$, $|a\rangle=|+-\rangle$, and $|b\rangle=|-+\rangle$. It can be shown from Eq.~\eqref{d21} that, in the case where the matrix elements $\rho_{ea}$, $\rho_{eb}$, $\rho_{ga}$, $\rho_{gb}$ are equal to zero
at the initial time, they remain equal to zero for any time. In this case,  
according to Tana\'{s} and Ficek \cite{Tanas}, the concurrence $C$ of the two-atom system is $C=\max(0,C_1,C_2)$, where $C_1=2(|\rho_{eg}|-\sqrt{\rho_{aa}\rho_{bb}})$ and $C_2=2(|\rho_{ab}|-\sqrt{\rho_{ee}\rho_{gg}})$.

We plot in Fig.~\ref{fig21} the time dependence of the concurrence $C$ for three different values of $z_{21}$.
We observe that the vacuum of the field can produce entanglement between the two atoms. Figures \ref{fig21}(a) and \ref{fig21}(b) show that,
when the atoms are close to each other, the magnitudes of the entanglement produced in the cases $|\psi(0)\rangle=|\psi_{1}\rangle$ (solid red lines) and $|\psi_{2}\rangle$ (dashed blue lines) are significant and almost equal to each other, and almost equal to that produced by atoms in free space (see the dotted black lines). The reason is that, when the separation between the atoms is small enough, the effect of radiation modes on the entanglement is dominant with respect to that of guided modes. We observe from Fig.~\ref{fig21}(c) that, when the separation between the atoms is large enough, the magnitudes of the entanglement produced in the cases $|\psi(0)\rangle=|\psi_{1}\rangle$ (solid red lines) and $|\psi_{2}\rangle$ (dashed blue lines) are small but not negligible, and differ significantly from each other and from the corresponding value that is produced by two atoms in free space (see the dotted black lines). We observe from Fig.~\ref{fig21} that, for the parameters used, the presence of the nanofiber reduces the peak value of the 
generated concurrence $C$. However, our additional calculations that are not shown here indicate that, depending on the parameters,
the presence of the nanofiber may reduce or increase the peak value of $C$ (see also Fig.~\ref{fig22}).

When the separation between the atoms is much larger than the wavelength of light, the effect of radiation modes on entanglement becomes negligible while the effect of guided modes survives. 
In order to illustrate the ability of the vacuum guided light field to produce entanglement between two atoms with a large separation, we plot in Fig.~\ref{fig22} the time dependence of the concurrence produced in the case where $z_{21}=100$ $\mu$m. We observe from the figure that,
 even though $z_{21}$ is very large as compared to the wavelength of light, the vacuum guided field can produce a finite entanglement. The  peak value of the produced concurrence (see the solid red and dashed blue lines) is substantially larger than the corresponding concurrence produced by the vacuum free-space field (see the dotted black line). Comparison between the solid red and dashed blue lines shows that the magnitude of the produced entanglement depends
on the excitation transfer direction specified by the ordering of the excited and un-excited atoms in the initial atomic states $|\psi_{1}\rangle$ and $|\psi_{2}\rangle$. Our results are consistent with the results of Ref.~ \cite{Gonzalez-Ballestero2015} for spontaneous generation of entanglement between two qubits chirally coupled to a one-dimensional waveguide.

\subsubsection{Symmetric and antisymmetric superposition states}

We now consider the cases where the initial state of the two-atom system is $|\psi(0)\rangle=|\psi_{\mathrm{sym}}\rangle$ or $|\psi_{\mathrm{asym}}\rangle$. Here, $|\psi_{\mathrm{sym}}\rangle=(|+-\rangle+e^{-i\varphi_{12}}|-+\rangle)/\sqrt2$ and $|\psi_{\mathrm{asym}}\rangle=(|+-\rangle-e^{-i\varphi_{12}}|-+\rangle)/\sqrt2$ are the symmetric and antisymmetric superposition states, with $\varphi_{12}$ being the phase of the cross-atom decay coefficient $\gamma_{12}$.

\begin{figure}[tbh]
\begin{center}
  \includegraphics{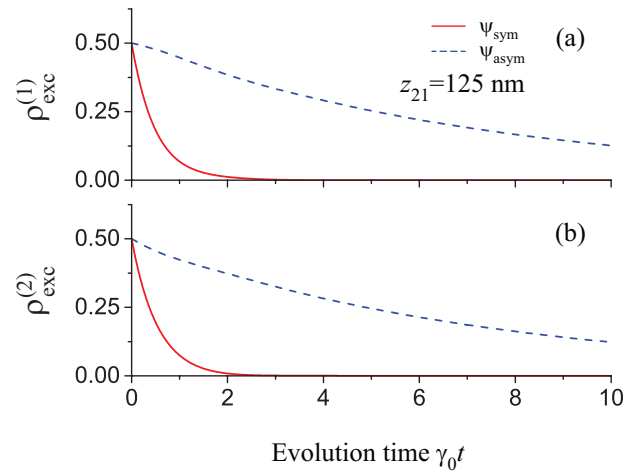}
 \end{center}
\caption{
Time evolution of the populations $\rho_{\mathrm{exc}}^{(1)}$ (a) and $\rho_{\mathrm{exc}}^{(2)}$ (b) of the upper levels of atoms 1 and 2, respectively,  in the cases where
the initial state of the two-atom system is $|\psi(0)\rangle=|\psi_{\mathrm{sym}}\rangle$ (solid red lines) or $|\psi_{\mathrm{asym}}\rangle$ (dashed blue lines).
The coordinates of the atoms are $r_1-a=r_2-a=200$ nm, $\varphi_1=\varphi_2=0$, and $z_2-z_1=125$ nm.
The dipole matrix elements of the atoms are $\mathbf{d}_1=\mathbf{d}_2=(d/\sqrt2)(i,0,-1)$, corresponding to the $\sigma_+$-polarized transitions with respect to the $y$ quantization axis. Other parameters are as for Figs.~\ref{fig2} and~\ref{fig4}.
}
\label{fig23}
\end{figure}

We plot in Fig.~\ref{fig23} the excited-state populations $\rho_{\mathrm{exc}}^{(1)}$ and $\rho_{\mathrm{exc}}^{(2)}$ of atoms 1 and 2, respectively, calculated for the cases where the initial state of the two-atom system is $|\psi(0)\rangle=|\psi_{\mathrm{sym}}\rangle$ (solid red lines) or $|\psi_{\mathrm{asym}}\rangle$ (dashed blue lines). The two atoms are aligned along the fiber axis with the separation $z_{21}=z_2-z_1=125$ nm. 
We observe from the figure that the decay of the excited-level populations of the atoms in the case of the initial state $|\psi_{\mathrm{sym}}\rangle$ (solid red lines) is much faster than that in the case of the initial state $|\psi_{\mathrm{asym}}\rangle$ (dashed blue lines). 
Comparison between Figs.~\ref{fig23}(a) and \ref{fig23}(b) shows that the decay of $\rho_{\mathrm{exc}}^{(1)}$  is almost the same as the decay of $\rho_{\mathrm{exc}}^{(2)}$ in the both cases.

\begin{figure}[tbh]
\begin{center}
  \includegraphics{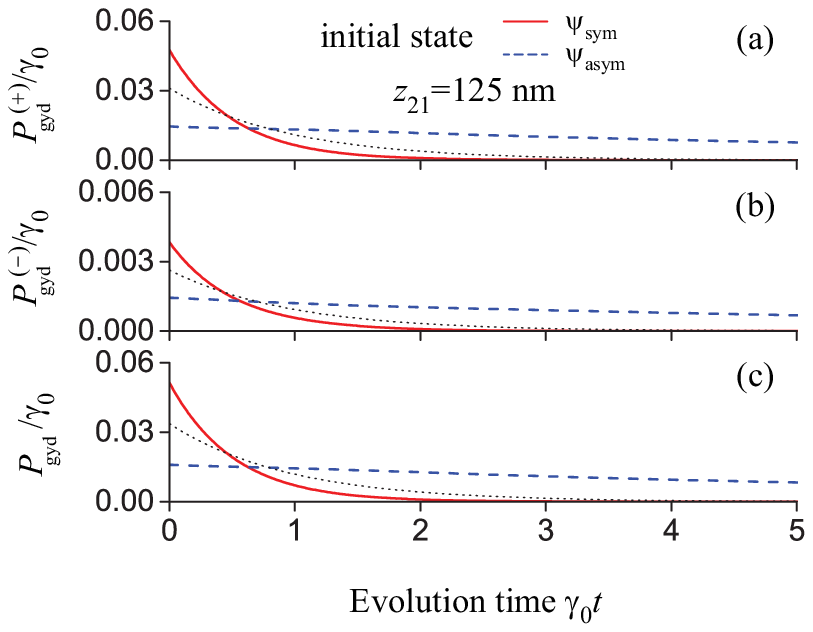}
 \end{center}
\caption{
Time evolution of the photon fluxes $P_{\mathrm{gyd}}^{(+)}$ (a), $P_{\mathrm{gyd}}^{(-)}$ (b), and $P_{\mathrm{gyd}}$ (c) in the cases where
the initial state of the two-atom system is $|\psi(0)\rangle=|\psi_{\mathrm{sym}}\rangle$ (solid red lines) or  $|\psi_{\mathrm{asym}}\rangle$ (dashed blue lines). Other parameters are as for Figs.~\ref{fig2}, \ref{fig4}, and \ref{fig23}.
The fluxes are normalized to the decay rate $\gamma_0$ of a single atom in free space.
The dotted black lines are for the case of a single excited atom.}
\label{fig24}
\end{figure}

We plot in Fig.~\ref{fig24} the time evolution of the fluxes $P_{\mathrm{gyd}}^{(+)}$ and $P_{\mathrm{gyd}}^{(-)}$ of photons emitted into guided modes in the positive and negative directions of the fiber axis, respectively, and the total guided-photon flux $P_{\mathrm{gyd}}$, calculated for the cases where the initial state of the two-atom system is $|\psi(0)\rangle=|\psi_{\mathrm{sym}}\rangle$ (solid red lines) or $|\psi_{\mathrm{asym}}\rangle$ (dashed blue lines). 
We observe that the photon flux $P_{\mathrm{gyd}}^{(+)}$ for the positive direction [see Fig.~\ref{fig24}(a)] is about one order larger than the photon flux $P_{\mathrm{gyd}}^{(-)}$ for the negative direction [see Fig.~\ref{fig24}(b)]. We also observe that the photon fluxes for the initial states $|\psi_{\mathrm{sym}}\rangle$ (solid red lines) and $|\psi_{\mathrm{asym}}\rangle$ (dashed blue lines) are different from each other. At the onset of the evolution, the photon fluxes in the cases of $|\psi_{\mathrm{sym}}\rangle$ (see solid red lines) and $|\psi_{\mathrm{asym}}\rangle$  (see dashed blue lines) are respectively larger and smaller than the photon fluxes in the case of a single excited atom (see the dotted black lines).
When the time is large enough, the opposite relationships hold true.
Thus, the states $|\psi_{\mathrm{sym}}\rangle$  and $|\psi_{\mathrm{asym}}\rangle$ correspond to superradiant and subradiant states \cite{Dicke, Haroche, Agarwal book} for guided modes in the case considered.

\begin{figure}[tbh]
\begin{center}
  \includegraphics{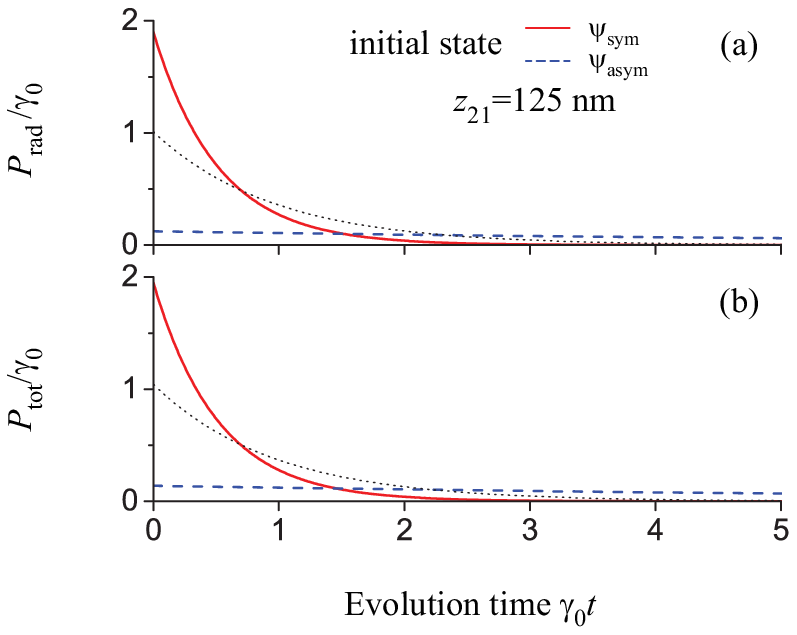}
 \end{center}
\caption{
Time evolution of the photon fluxes $P_{\mathrm{rad}}$ (a) and $P_{\mathrm{tot}}$ (b) in the cases where
the initial state of the two-atom system is $|\psi(0)\rangle=|\psi_{\mathrm{sym}}\rangle$ (solid red lines) or $|\psi_{\mathrm{asym}}\rangle$ (dashed blue lines).
Other parameters are as for Figs.~\ref{fig2}, \ref{fig4}, and \ref{fig23}.
The fluxes are normalized to the decay rate $\gamma_0$ of a single atom in free space.
The dotted black lines are for the case of a single excited atom.}
\label{fig25}
\end{figure}

We plot in Fig.~\ref{fig25} the time evolution of the flux $P_{\mathrm{rad}}$ of photons emitted into radiation modes and the total photon flux $P_{\mathrm{tot}}$ for the parameters of Fig.~\ref{fig24}. We observe that the fluxes $P_{\mathrm{rad}}$ and $P_{\mathrm{tot}}$ in the case
of $|\psi(0)\rangle=|\psi_{\mathrm{sym}}\rangle$ (solid red lines) are different from those in the case of $|\psi_{\mathrm{asym}}\rangle$ (dashed blue lines). When the interaction time is short enough, the values of $P_{\mathrm{rad}}$ and $P_{\mathrm{tot}}$ in the case of $|\psi(0)\rangle=|\psi_{\mathrm{sym}}\rangle$ (solid red lines) are larger than the corresponding values in the case of a single excited atom (dotted black lines). 
Meanwhile, when the interaction time is not too long, the values of $P_{\mathrm{rad}}$ and $P_{\mathrm{tot}}$ in the case of $|\psi(0)\rangle=|\psi_{\mathrm{asym}}\rangle$ (dashed blue lines) are smaller than the corresponding values in the case of a single excited atom (dotted black lines). Thus, the superposition states $|\psi_{\mathrm{sym}}\rangle$ and $|\psi_{\mathrm{asym}}\rangle$ are also superradiant and subradiant states \cite{Dicke, Haroche, Agarwal book} for radiation modes in the case considered.

\begin{figure}[tbh]
\begin{center}
  \includegraphics{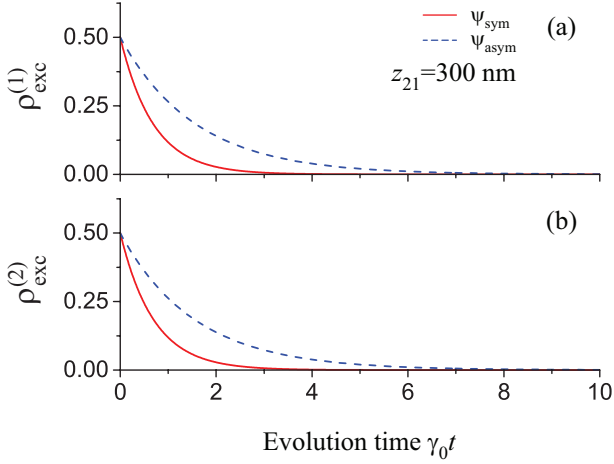}
 \end{center}
\caption{
Same as Fig.~\ref{fig23} but for $z_{21}=300$ nm.
}
\label{fig26}
\end{figure}

\begin{figure}[tbh]
\begin{center}
  \includegraphics{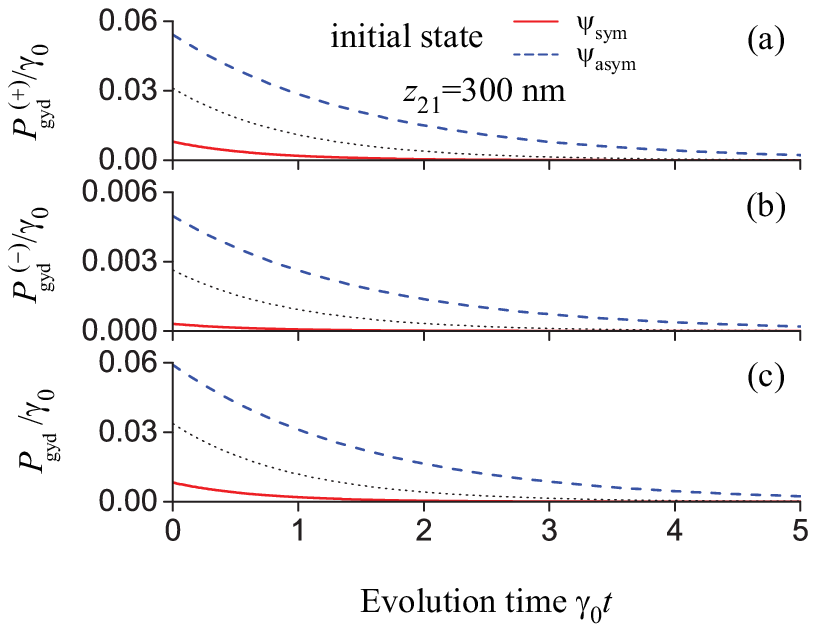}
 \end{center}
\caption{
Same as Fig.~\ref{fig24} but for $z_{21}=300$ nm.
}
\label{fig27}
\end{figure}

\begin{figure}[tbh]
\begin{center}
  \includegraphics{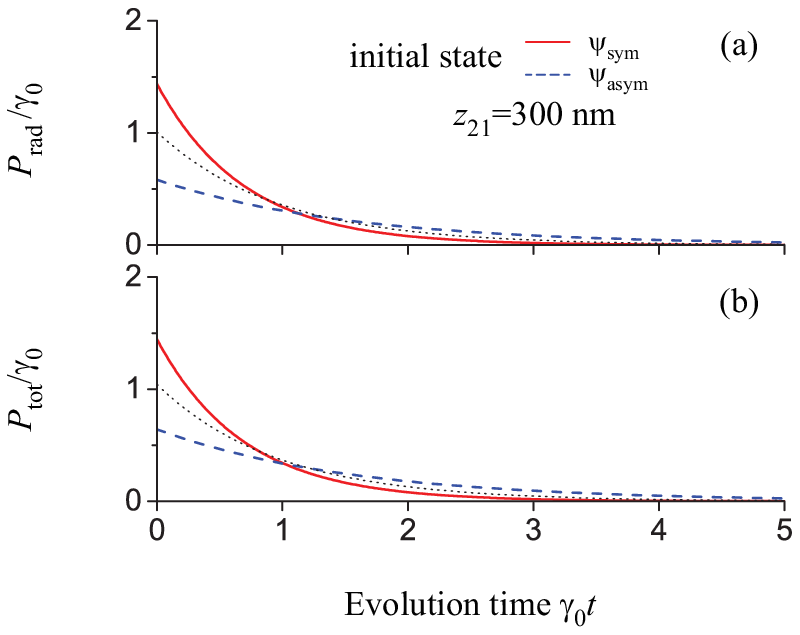}
 \end{center}
\caption{
Same as Fig.~\ref{fig25} but for $z_{21}=300$ nm.
}
\label{fig28}
\end{figure}

It is interesting to note that an atomic superposition state can be a superradiant state for radiation modes but a subradiant state for guided modes. In order to illustrate such a situation, we plot in Figs.~\ref{fig26}--\ref{fig28} the results of calculations for the case where $z_{21}=300$ nm. Figure \ref{fig26} shows that the decay of the excited-level populations in the case of $|\psi_{\mathrm{sym}}\rangle$ (solid red lines) is faster than
that in the case of $|\psi_{\mathrm{asym}}\rangle$ (dashed blue lines). Meanwhile, according to Fig.~\ref{fig27}, the fluxes of photon emitted into radiation modes in the cases of the initial states $|\psi_{\mathrm{sym}}\rangle$ (solid red lines) and $|\psi_{\mathrm{asym}}\rangle$ (dashed blue lines) are respectively weaker and stronger than those in the case of a single excited atom (dotted black lines). Thus, the superposition states $|\psi_{\mathrm{sym}}\rangle$ and $|\psi_{\mathrm{asym}}\rangle$ are respectively subradiant and superradiant states for emission into guided modes. The result of Figs.~\ref{fig26} and \ref{fig27} do no contradict the energy conservation law. Indeed, as already mentioned above, in addition to emission into guided modes, there is emission into radiation modes.
According to Fig.~\ref{fig28}, the fluxes of photon emitted into guided modes in the cases of the initial states $|\psi_{\mathrm{sym}}\rangle$ (solid red lines) and $|\psi_{\mathrm{asym}}\rangle$ (dashed blue lines) are respectively stronger and weaker than those in the case of a single excited atom (dotted black lines). This result means that the superposition states $|\psi_{\mathrm{sym}}\rangle$ and $|\psi_{\mathrm{asym}}\rangle$ are respectively superradiant and subradiant states for emission into radiation modes as well as for the total emission into both types of modes. These collective effects are opposite to those collective effects occurring in emission into guided modes. The difference is caused by the action of cross-atom interference on the emission rate.

\begin{figure}[tbh]
\begin{center}
  \includegraphics{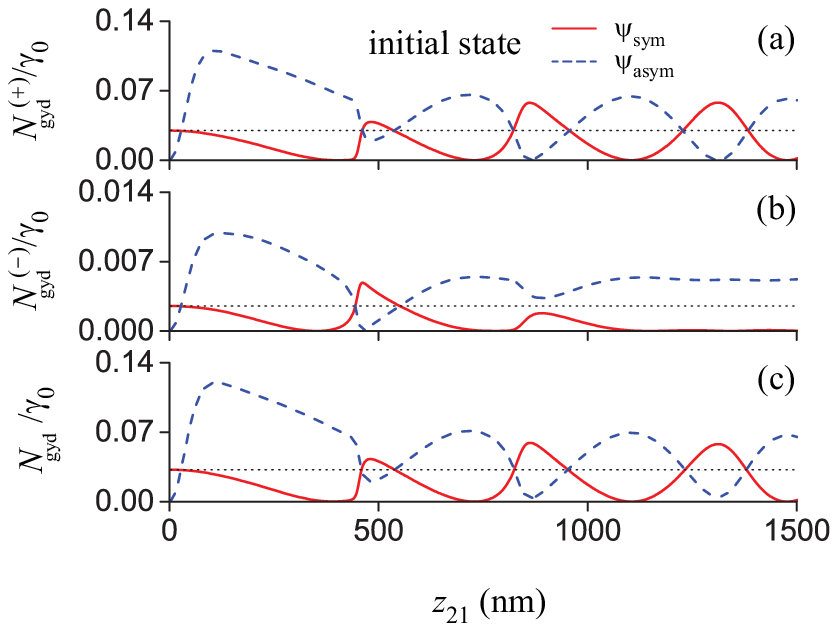}
 \end{center}
\caption{
Dependences of the mean emitted photon numbers $N_{\mathrm{gyd}}^{(+)}$ (a), $N_{\mathrm{gyd}}^{(-)}$ (b), and $N_{\mathrm{gyd}}$ (c) on the axial separation $z_{21}$ between the atoms in the cases where
the initial state of the two-atom system is $|\psi(0)\rangle=|\psi_{\mathrm{sym}}\rangle$ (solid red lines) or  $|\psi_{\mathrm{asym}}\rangle$ (dashed blue lines). The radial and azimuthal coordinates of the atoms are $r_1-a=r_2-a=200$ nm and $\varphi_1=\varphi_2=0$, respectively.
Other parameters are as for Figs.~\ref{fig2}, \ref{fig4}, and \ref{fig23}.
The dotted black lines are for the case of a single excited atom.}
\label{fig29}
\end{figure}

\begin{figure}[tbh]
\begin{center}
  \includegraphics{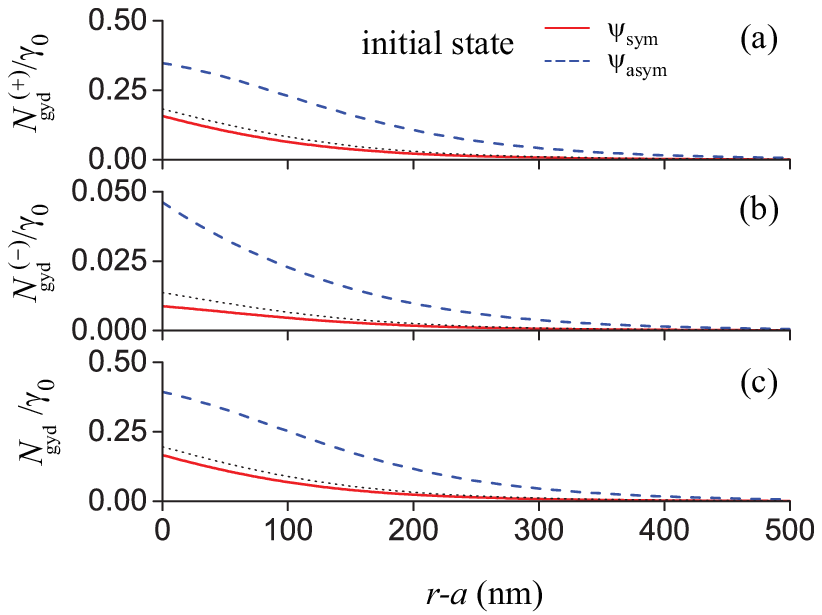}
 \end{center}
\caption{
Dependences of the mean emitted photon numbers $N_{\mathrm{gyd}}^{(+)}$ (a), $N_{\mathrm{gyd}}^{(-)}$ (b), and $N_{\mathrm{gyd}}$ (c) on the distance $r-a$ from the atoms to the fiber surface in the cases where
the initial state of the two-atom system is $|\psi(0)\rangle=|\psi_{\mathrm{sym}}\rangle$ (solid red lines) or  $|\psi_{\mathrm{asym}}\rangle$ (dashed blue lines). The coordinates of the atoms are $r_1=r_2=r$, $\varphi_1=\varphi_2=0$, and $z_2-z_1=150$ nm.
Other parameters are as for Figs.~\ref{fig2}, \ref{fig4}, and \ref{fig23}.
The dotted lines are for the case of a single excited atom.}
\label{fig30}
\end{figure}

We plot in Figs.~\ref{fig29} and \ref{fig30} the dependences of the mean emitted guided photon numbers on the axial atomic separation $z_{21}$ and the atom-to-surface distance $r-a$, respectively. We observe from Figs.~\ref{fig29} and \ref{fig30} that the mean photon number $N_{\mathrm{gyd}}^{(+)}$ for the positive direction is about one order larger than the mean photon number $N_{\mathrm{gyd}}^{(-)}$ for the negative direction. It is clear from the figures that the mean emitted guided photon number $N_{\mathrm{gyd}}$ and its directional components $N_{\mathrm{gyd}}^{(+)}$ and $N_{\mathrm{gyd}}^{(-)}$ depend on the axial atomic separation $z_{21}$, the atom-surface distance $r-a$, and the initial superposition state. When we compare the solid red and dashed blue lines of Fig.~\ref{fig29} with the dotted lines of this figure, we see that, depending on the axial atomic separation $z_{21}$ and the initial superposition state, the probability of emitting a photon into guided modes may be enhanced or suppressed. We observe from Fig.~\ref{fig29} that, depending on $z_{21}$, the values of $N_{\mathrm{gyd}}^{(+)}$, $N_{\mathrm{gyd}}^{(-)}$, and $N_{\mathrm{gyd}}$ in the case of the initial state $|\psi(0)\rangle=|\psi_{\mathrm{sym}}\rangle$ (solid red lines) may be larger or smaller than the corresponding values in the case of the initial state $|\psi(0)\rangle=|\psi_{\mathrm{asym}}\rangle$ (dashed blue lines). 
We observe from Fig.~\ref{fig29} that there exist regions of $z_{21}$ where $N_{\mathrm{gyd}}$ and its directional components $N_{\mathrm{gyd}}^{(+)}$ and $N_{\mathrm{gyd}}^{(-)}$ for the two-atom case (see the solid red and dashed blue lines) are several times larger than the corresponding values for a single excited atom (see the dotted black lines). And there also exist regions of $z_{21}$ where $N_{\mathrm{gyd}}$ and its directional components $N_{\mathrm{gyd}}^{(+)}$ and $N_{\mathrm{gyd}}^{(-)}$ for the two-atom case are almost zero. 
These features are signatures of the collective effect in spontaneous emission into guided modes. 
We emphasize again that an increase or a decrease in the mean number of photons emitted into guided modes is associated with a decrease or an increase, respectively, in the mean number of photons emitted into radiation modes.

\section{Summary}
\label{sec:summary}
In this paper, we have studied the coupling between two two-level atoms with arbitrarily polarized dipoles in the vicinity of a nanofiber. 
We have derived the master equation for the atoms interacting with the vacuum of the field in the guided and radiation modes of the nanofiber. 
We have obtained the expressions for the single-atom and cross-atom decay coefficients and their directional components. We have also got the expression for the dipole-dipole interaction coefficients. We have studied numerically the case where the atomic dipoles are circularly polarized and, consequently, the rate of emission depends on the propagation direction and the radiative interaction between the atoms is chiral. We have examined the time evolution of the atoms for different initial one-excitation states. We have calculated the fluxes and mean numbers of photons spontaneously emitted into guided modes in the positive and negative directions of the fiber axis. We have shown that the chiral radiative coupling modifies the collective emission of the atoms. We have observed that the modifications strongly depend on the initial state of the atomic system, the radiative transfer direction, the distance between the atoms, and the distance from the atoms to the fiber surface. 

\begin{acknowledgments}
We thank Th. Busch for helpful comments and discussions.
F.L.K. acknowledges support for this work from the Okinawa Institute of Science and Technology Graduate University.
\end{acknowledgments}


\appendix

\section{Guided modes of a nanofiber}
\label{sec:guided}

Consider a nanofiber that is a silica cylinder of radius $a$ and refractive index $n_1$ and is surrounded by an infinite background medium of refractive index $n_2$,
where $n_2<n_1$. The radius of the nanofiber is well below a given free-space wavelength $\lambda$ of light. Therefore, the nanofiber supports only the hybrid fundamental modes HE$_{11}$ corresponding to the given wavelength $\lambda$ \cite{fiber books}. The light field in such a mode is strongly guided. It penetrates into the outside of the nanofiber in the form of an evanescent wave carrying a significant fraction of energy \cite{fibermode}.
For a fundamental guided mode HE$_{11}$ of a light field of frequency $\omega$ (free-space wavelength $\lambda=2\pi c/\omega$ and free-space wave number $k=\omega/c$), the propagation constant $\beta$ is determined by the
fiber eigenvalue equation \cite{fiber books}
\begin{eqnarray}\label{g1}
\frac{J_0(h a)}{h a J_1(h a)}&=&
-\frac{n_1^2+n_2^2}{2n_1^2}\frac{K_1'(q a)}{q a K_1(q a)}+ \frac{1}{h^2 a^2}
\nonumber\\&&\mbox{}
-\Bigg[\left(\frac{n_1^2-n_2^2}{2n_1^2}\frac{K_1'(q a)}{q a K_1(q a)}\right)^2
\nonumber\\&&\mbox{}
+\frac{\beta^2}{n_1^2 k^2}\left(\frac{1}{q^2a^2}+\frac{1}{h^2a^2}\right)^2 \Bigg]^{1/2}.
\end{eqnarray}
Here the parameters $h=(n_1^2k^2-\beta^2)^{1/2}$ and $q=(\beta^2-n_2^2k^2)^{1/2}$ characterize the fields inside and outside the fiber, respectively. The notations $J_n$ and $K_n$ stand for the Bessel functions of the first kind and the modified Bessel functions of the second kind, respectively. 

According to \cite{fiber books}, the cylindrical-coordinate vector components of the profile function $\mathbf{e}(\mathbf{r})$ 
of the electric part of the fundamental guided mode that propagates in the forward ($+\hat{\mathbf{z}}$) direction and is
counterclockwise quasicircularly polarized are given, for $r<a$, by
\begin{eqnarray}\label{g2}
e_{r}&=&iC\frac{q}{h}\frac{K_1(qa)}{J_1(ha)}[(1-s)J_0(hr)-(1+s)J_2(hr) ],
\nonumber\\
e_{\varphi}&=&-C\frac{q}{h}\frac{K_1(qa)}{J_1(ha)}[(1-s)J_0(hr)+(1+s)J_2(hr) ],
\nonumber\\
e_{z}&=&C\frac{2q}{\beta}\frac{K_1(qa)}{J_1(ha)}J_1(hr),
\end{eqnarray}
and, for $r>a$, by
\begin{eqnarray}\label{g3}
e_{r}&=&iC[(1-s)K_0(qr)+(1+s)K_2(qr) ],
\nonumber\\
e_{\varphi}&=&-C[(1-s)K_0(qr)-(1+s)K_2(qr) ],
\nonumber\\
e_{z}&=&C\frac{2q}{\beta}K_1(qr).
\end{eqnarray}
Here the parameter $s$ is defined as
\begin{equation}\label{g4} 
s=\frac{{1}/{h^2a^2}+{1}/{q^2a^2}}{{J_1^\prime (ha)}/{haJ_1(ha)}+{K_1^\prime (qa)}/{qaK_1(qa)}}.
\end{equation}
The parameter $C$ is the normalization coefficient. We take $C$ to be a positive real number and use the normalization condition 
\begin{equation}\label{g5}
\int _{0}^{2\pi}d\varphi\int _{0}^{\infty}n_{\mathrm{ref}}^2\,|\mathbf{e}|^2r\,dr=1.
\end{equation}
Here $n_{\mathrm{ref}}(r)=n_1$ for $r<a$, and $n_{\mathrm{ref}}(r)=n_2$ for $r>a$.
We note that the axial component $e_{z}$ is significant in the case of nanofibers \cite{fibermode}. This makes guided modes of nanofibers very different from plane-wave modes of the field in free space and from guided modes of conventional (weakly guiding) fibers \cite{fibermode,fiber books}.

We label quasicircularly polarized fundamental guided modes HE$_{11}$ by using a mode index $\mu=(\omega,f,l)$, where $\omega$ is the mode frequency, $f=+1$ or $-1$ (or simply $+$ or $-$) 
denotes the forward ($+\hat{\mathbf{z}}$) or backward ($-\hat{\mathbf{z}}$) propagation direction, respectively, and $l=+1$ or $-1$ (or simply $+$ or $-$) 
denotes the counterclockwise  or clockwise circulation, respectively, of the transverse component of the polarization around the axis $+\hat{\mathbf{z}}$. 
In the cylindrical coordinates, the components of the profile function $\mathbf{e}^{(\mu)}(\mathbf{r})$ of the electric part of the quasicircularly polarized fundamental guided mode $\mu$ are given by
\begin{eqnarray}\label{g6}
e_{r}^{(\mu)}&=&e_{r},
\nonumber\\
e_{\varphi}^{(\mu)}&=&le_{\varphi},
\nonumber\\
e_{z}^{(\mu)}&=& fe_{z}.
\end{eqnarray}
Consequently, the profile function of the quasicircularly polarized mode $(\omega, f, l)$ can be written as
\begin{eqnarray}\label{g7}
\mathbf{e}^{(\omega fl)}&=&\hat{\mathbf{r}}e^{(\omega fl)}_r+\hat{\boldsymbol{\varphi}}e^{(\omega fl)}_\varphi+\hat{\mathbf{z}}e^{(\omega fl)}_z
\nonumber\\ 
&=&\hat{\mathbf{r}}e_r+l\hat{\boldsymbol{\varphi}}e_\varphi+f\hat{\mathbf{z}}e_z,
\end{eqnarray}
where the notations 
$\hat{\mathbf{r}} = \hat{\mathbf{x}}\cos\varphi + \hat{\mathbf{y}}\sin\varphi$,  
$\hat{\boldsymbol{\varphi}} = -\hat{\mathbf{x}}\sin\varphi + \hat{\mathbf{y}}\cos\varphi$, 
and $\hat{\mathbf{z}}$ stand for the unit basis vectors of the cylindrical coordinate system $\{r,\varphi,z\}$.
Here $\hat{\mathbf{x}}$ and $\hat{\mathbf{y}}$ are the unit basis vectors of the Cartesian coordinate system for the fiber transverse plane $xy$.

We have the following symmetry relations: 
\begin{eqnarray}\label{A10}
e_r^{(\omega,f,l)}&=&e_r^{(\omega,-f,l)}=e_r^{(\omega,f,-l)},\nonumber\\
e_{\varphi}^{(\omega,f,l)}&=&e_{\varphi}^{(\omega,-f,l)}=-e_{\varphi}^{(\omega,f,-l)},\nonumber\\
e_z^{(\omega,f,l)}&=&-e_z^{(\omega,-f,l)}=e_z^{(\omega,f,-l)},
\end{eqnarray}
and
\begin{equation}\label{A11}
e_r^{(\mu)*}=-e_r^{(\mu)},\quad
e_\varphi^{(\mu)*}=e_\varphi^{(\mu)},\quad
e_z^{(\mu)*}=e_z^{(\mu)}.
\end{equation}

\section{Radiation modes of a nanofiber}
\label{sec:radiation}

For the radiation modes, we have $-kn_2<\beta<kn_2$.
The characteristic parameters for the field in the inside and outside of the fiber are $h=\sqrt{k^2n_1^2-\beta^2}$ and $q=\sqrt{k^2n_2^2-\beta^2}$, respectively.
The mode functions of the electric parts of the radiation modes $\nu=(\omega\beta m l)$
\cite{fiber books} are given, for $r<a$, by
\begin{eqnarray}\label{q1}
e_r^{(\nu)}&=&
\frac{i}{h^2}\left[\beta hAJ'_m(hr)+im\frac{\omega\mu_0}{r}BJ_m(hr)\right],\nonumber\\ 
e_{\varphi}^{(\nu)}&=&
\frac{i}{h^2}\left[im\frac{\beta}{r}AJ_m(hr)-h\omega\mu_0BJ'_m(hr)\right],\nonumber\\
e_z^{(\nu)}&=&AJ_m(hr),
\end{eqnarray}
and, for $r>a$, by 
\begin{eqnarray}\label{q2}
e_r^{(\nu)}&=&
\frac{i}{q^2}\sum_{j=1,2}
\left[\beta q C_jH^{(j)\prime}_m(qr)+im\frac{\omega\mu_0}{r}D_jH^{(j)}_m(qr)\right],\nonumber\\
e_{\varphi}^{(\nu)}&=&
\frac{i}{q^2}\sum_{j=1,2}
\left[im\frac{\beta}{r}C_jH^{(j)}_m(qr)-q\omega\mu_0D_jH^{(j)\prime}_m(qr)\right], \nonumber\\
e_z^{(\nu)}&=&\sum_{j=1,2}C_jH_m^{(j)}(qr).
\end{eqnarray}
Here $A$ and $B$ as well as $C_j$ and $D_j$ with $j=1,2$ are coefficients.
The coefficients $C_j$ and $D_j$ are related to the coefficients $A$ and $B$ as 
\cite{Tromborg}
\begin{eqnarray}\label{q3}
C_j&=&(-1)^{j}\frac{i\pi q^2a}{4n_2^2}(AL_j+i\mu_0cBV_j),\nonumber\\
D_j&=&(-1)^{j-1}\frac{i\pi q^2a}{4}(i\epsilon_0cAV_j-BM_j),
\end{eqnarray}
where
\begin{eqnarray}\label{q4}
V_j&=&\frac{mk\beta}{ah^2q^2}
(n_2^2-n_1^2)
J_m(ha)H_m^{(j)*}(qa),\nonumber\\
M_j&=&\frac{1}{h}J'_m(ha)H_m^{(j)*}(qa)
-\frac{1}{q}J_m(ha)H_m^{(j)*\prime}(qa),\nonumber\\
L_j&=&\frac{n_1^2}{h}J'_m(ha)H_m^{(j)*}(qa)
-\frac{n_2^2}{q}J_m(ha)H_m^{(j)*\prime}(qa).\nonumber\\
\end{eqnarray}
We specify two polarizations by choosing $B=i\eta A$ and $B=-i\eta A$ for $l=+$
and $l=-$, respectively. We take $A$ to be a real number.
The orthogonality of the modes requires
\begin{eqnarray}\label{q5}
&&\int _0^{2\pi}d\varphi\int _{0}^{\infty}n_{\mathrm{ref}}^2
\left[\mathbf{e}^{(\nu)}\mathbf{e}^{(\nu')*}\right]_{\beta=\beta',m=m'}
\;rdr \nonumber\\&&
=N_{\nu}\delta_{ll'}\delta(\omega-\omega').
\end{eqnarray}
This leads to
\begin{equation}\label{q6}
\eta=\epsilon_0c\sqrt{\frac{n_2^2|V_j|^2+|L_j|^2}{|V_j|^2+n_2^2|M_j|^2}}.
\end{equation}
The constant $N_{\nu}$ is given by 
\begin{equation}\label{q7}
N_{\nu}=\frac{8\pi \omega}{q^2}\left(n_2^2|C_j|^2+\frac{\mu_0}{\epsilon_0}|D_j|^2\right).
\end{equation}
We use the normalization $N_{\nu}=1$. 

We have the following symmetry relations: 
\begin{eqnarray}\label{q8}
e_r^{(\omega,\beta, m,l)}&=&-e_r^{(\omega,-\beta, m,-l)},\nonumber\\
e_{\varphi}^{(\omega,\beta, m,l)}&=&-e_{\varphi}^{(\omega,-\beta, m,-l)},\nonumber\\
e_z^{(\omega,\beta, m,l)}&=&e_z^{(\omega,-\beta, m,-l)},\nonumber\\
\end{eqnarray}
\begin{eqnarray}\label{q9}
e_{r}^{(\omega,\beta, m,l)}&=&(-1)^m e_{r}^{(\omega,\beta, -m,-l)},\nonumber\\
e_{\varphi}^{(\omega,\beta, m,l)}&=&(-1)^{m+1} e_{\varphi}^{(\omega,\beta, -m,-l)},\nonumber\\
e_{z}^{(\omega,\beta, m,l)}&=&(-1)^m e_{z}^{(\omega,\beta, -m,-l)},
\end{eqnarray} 
and
\begin{equation}\label{q10}
e_r^{(\nu)*}=-e_r^{(\nu)},\quad
e_\varphi^{(\nu)*}=e_\varphi^{(\nu)},\quad
e_z^{(\nu)*}=e_z^{(\nu)}.
\end{equation}

\end{document}